\documentclass[Journal]{IEEEtran}
\IEEEoverridecommandlockouts

\usepackage{cite}
\usepackage[hidelinks]{hyperref}
\usepackage{amsmath,amssymb,amsfonts}
\usepackage{algorithm}
\usepackage{algorithmic}
\usepackage{graphicx}
\usepackage{booktabs}
\usepackage{bm}
\usepackage{textcomp}
\usepackage{xcolor}
\usepackage{epstopdf}
\usepackage{caption}

\usepackage{color}
\usepackage{multirow}
\usepackage{array}
\usepackage{amsthm}
\usepackage{stfloats}
\usepackage{subfigure}
\usepackage{gensymb}
\usepackage{url}
\usepackage{verbatim}
\usepackage{pifont}
\usepackage{makecell}
\usepackage{flushend}
\definecolor{color_radar}{rgb}{0.65, 0.85, 0.35}
\definecolor{color_comm}{rgb}{0.2, 0.5, 0.9}

\usepackage{hyperref}
\usepackage{cleveref}
\hypersetup{colorlinks = true,
    linkcolor = black,
    urlcolor = blue,
    citecolor = black}

\allowdisplaybreaks[4]
\begin{document}
\title{Cooperative OFDM-ISAC Networks: Performance Analysis and Resource Allocation\\
\thanks{S. Zhang and M. Li are with the School of Information and Communication Engineering, Dalian University of Technology, Dalian 116024, China (e-mail: zhangshoushuo@mail.dlut.edu.cn; mli@dlut.edu.cn).}
\thanks{R. Liu is with the Institute for Digital Communications (IDC), Friedrich-Alexander-University Erlangen-Nuremberg (FAU), 91058 Erlangen, Germany (e-mail: rang.liu@fau.de).}
\thanks{Q. Liu is with the School of Computer Science and Technology, Dalian University of Technology, Dalian 116024, China (e-mail: qianliu@dlut.edu.cn).}
}
\author{Shoushuo~Zhang,
        Rang~Liu,~\IEEEmembership{Member,~IEEE,}
        Qian~Liu,~\IEEEmembership{Member,~IEEE,}        and~Ming~Li,~\IEEEmembership{Senior~Member,~IEEE}
}
\maketitle
\pagestyle{empty}
\thispagestyle{empty}
\begin{abstract}
Cooperative integrated sensing and communication (ISAC) based on orthogonal frequency-division multiplexing (OFDM) enables network-wide sensing by exploiting the spatial diversity of multi-base-station (BS). This paper studies performance analysis and time-frequency resource allocation for a multi-BS cooperative OFDM-ISAC network with fine-grained resource-element (RE)-level orthogonal coordination. Two fusion architectures are considered: signal-level fusion (SLF), which forwards raw echoes to a fusion center, and parameter-level fusion (PLF), which reports only local delay/Doppler estimates and their uncertainty information. For SLF, we derive the Cram\'er--Rao bound (CRB) for joint target position and velocity estimation. For PLF, we develop a two-stage CRB-like metric by combining local delay/Doppler uncertainty characterization with first-order geometric error propagation, and show that only an oracle ML-based PLF benchmark can asymptotically attain the SLF CRB under restrictive conditions. Based on these results, we formulate a joint RE-selection and power-allocation problem under network-wide RE exclusivity, per-BS power budgets, a communication sum-rate constraint, and a sidelobe-amplitude constraint on the delay-Doppler ambiguity function. An efficient solution is developed via Schur-complement reformulations and penalty-based alternating optimization. Numerical results validate the analysis, demonstrate effective ambiguity-sidelobe suppression and consistent localization/velocity gains over representative baselines, while revealing geometry-dependent SLF-PLF performance gaps.
\end{abstract}

\section{Introduction}

Integrated sensing and communication (ISAC) enables wireless infrastructure to simultaneously provide connectivity and environmental perception, making it a key technology for future networks \cite{Zhen_Du_WCM_2025, Rang_Liu_WCM_2023, Rang_Liu_Proc_2026}.
Compared with sensing using a single base station (BS), cooperative ISAC exploits spatial diversity and multi-view observations across distributed BSs to enlarge sensing coverage and improve robustness against blockage, multipath, and transmit leakage \cite{Kaitao_Meng_WCM_2025, Jun_Tang_TWC_2025, Zhiqing_Wei_Net_2024, Xiaoyun_Wang_WCM_2025, HaoJin_Li_WCM_2025, Rang_Liu_WCM_2026}. These advantages are particularly attractive in OFDM systems, where the time--frequency grid offers fine-grained flexibility for multiplexing sensing and communication functions.

Despite this promise, cooperative OFDM-ISAC still faces two tightly coupled issues. The first is transmitter-side resource allocation. Since sensing pilots and communication symbols share the same OFDM grid across multiple BSs, the network must decide how to schedule time-frequency resources and allocate transmit power while maintaining communication throughput and sensing reliability. The second issue is receiver-side performance characterization. Different fusion architectures preserve different amounts of sensing information: signal-level fusion (SLF) can exploit raw observations at the cost of heavy fronthaul overhead, whereas parameter-level fusion (PLF) exchanges only low-dimensional local estimates and is therefore more scalable, but may incur non-negligible information loss. These two aspects are inherently connected, because meaningful transmitter-side design should be evaluated according to post-fusion sensing performance rather than local per-link heuristics.

\begin{figure*}[!t]
 \centering
 \vspace{-0.0 cm}
 \includegraphics[width=7.0in]{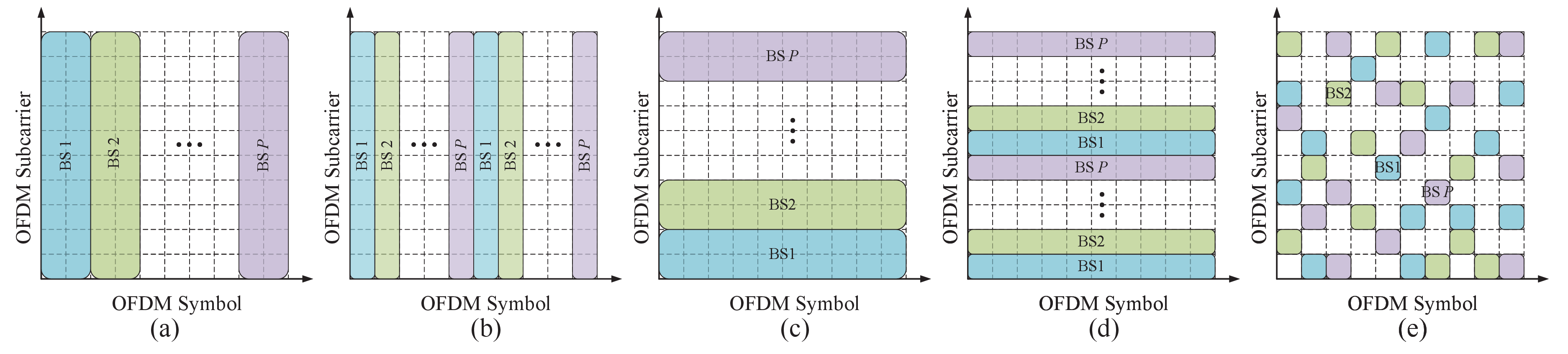}
 \caption{Resource allocation schemes for cooperative sensing: (a) time-division block (\textbf{TDB}), (b) time-division interleaved (\textbf{TDI}), (c) frequency-division block (\textbf{FDB}), (d) frequency-division interleaved (\textbf{FDI}), and (e) the proposed non-periodic 2D allocation.}
 \label{fig:Re allocation}
 \vspace{-0.2 cm}
\end{figure*}

For the transmitter-side issue, this paper focuses on network-wide resource element (RE)-level orthogonal coordination, where each sensing RE is assigned to at most one Tx-BS. This suppresses inter-BS interference and simplifies Tx-Rx association, making multistatic processing tractable. However, orthogonalization fragments the sensing aperture. As illustrated in Fig.~\ref{fig:Re allocation}(a)--(d) and Table~\ref{tab:ofdm_RE allocation_compare}, conventional one-dimensional (1D) orthogonal schemes suffer from a resolution--ambiguity tension. Specifically, contiguous partitioning in time or frequency, i.e., time-division block (TDB) \cite{Xiaoyv_Yang_TWC_2024} and frequency-division block (FDB) \cite{Xiaoyv_Yang_TWC_2025}, avoids periodic-sampling ambiguity replicas but reduces the effective coherent processing interval (CPI) or bandwidth, thereby degrading Doppler/velocity or range resolution, respectively. In contrast, periodic interleaving, i.e., time-division interleaved (TDI) and frequency-division interleaved (FDI) \cite{Kawon_Han_TWC_2026}, preserves the corresponding span on average but introduces Doppler or range ambiguities due to periodic sampling. These observations motivate adaptive, non-periodic two-dimensional (2D) allocation on the OFDM grid to jointly shape the effective bandwidth and CPI while controlling ambiguity sidelobes.

Two-dimensional time--frequency scheduling is well aligned with modern cellular standards and has been extensively studied for \emph{single-BS monostatic} OFDM-ISAC. Existing works optimize sparse RE selection and power allocation using Cram\'er--Rao bound (CRB)-type sensing objectives under communication constraints \cite{Keskin_TSP_2021, Iqbal_APWiMob_2025}, and further incorporate sensing--communication prioritization and peak sidelobe level (PSL)-based ambiguity control \cite{ZhangFan_JSEC_2024, Peishi_Li_TSP_2025}. In cooperative \emph{multi-BS multistatic} networks, however, prior studies have mainly emphasized transmit beamforming, BS activation, and fronthaul-aware processing \cite{Wenrui_Li_TWC_2025, Jie_Chen_TCOM_2025, Sifan_Liu_WCNC_2024, Zihuan_Wang_JSAC_2025}, sparse waveform design for ambiguity/interference mitigation \cite{Amani_2025}, or localization algorithms and performance bounds \cite{ZhiqingWei_TVT_2024, ZiJie_Wang_JSEC, Pucci_SPAWC_2024}. However, a unified framework that simultaneously characterizes post-fusion sensing performance under different fusion architectures and develops time-frequency resource allocation under network-wide RE exclusivity, communication coexistence, and ambiguity control is still lacking.

This gap is closely tied to the choice of fusion architecture on the receiver-side. SLF forwards raw echoes from the receivers to a fusion center and can, in principle, preserve the full Fisher information \cite{Daniel_Hack_TSP_2014, Pucci_WCL_2025}, but its fronthaul payload scales with the number of collected samples. In contrast, PLF forwards only low-dimensional local summaries, such as delay/Doppler estimates and their uncertainty information, thereby substantially reducing the fronthaul burden \cite{Qin_Shi_JSEC_2022, Mao_TSP_2024}, at the cost of architecture- and geometry-dependent information loss. Although algorithms have been developed for both paradigms \cite{ZhiqingWei_TVT_2024, ZiJie_Wang_JSEC}, a unified and tractable framework for cooperative OFDM-ISAC networks is still lacking. Such a framework should not only characterize the post-fusion sensing performance of SLF and PLF in a comparable manner, thereby distinguishing information-preserving and information-lossy fusion regimes and clarifying how closely PLF can approach SLF, but also enable transmitter-side time-frequency resource-allocation design based on end-to-end sensing performance.

\begin{table}[t]
\centering
\caption{Qualitative comparison of cooperative OFDM sensing resource-allocation schemes.}
\label{tab:ofdm_RE allocation_compare}
\renewcommand{\arraystretch}{1.2}
\setlength{\tabcolsep}{6pt}
\begin{tabular}{|c|cc|cc|}
\hline
\multirow{2}{*}{\thead{Scheme}}
& \multicolumn{2}{c|}{\thead{Range}}
& \multicolumn{2}{c|}{\thead{Velocity}} \\
\cline{2-5}
& \thead{Resolution }
& \thead{Ambiguity}
& \thead{Resolution}
& \thead{Ambiguity} \\
\hline
\textbf{TDB}      & High & No  & Low  & No  \\
\textbf{TDI}      & High & No  & High & Yes \\
\textbf{FDB}      & Low  & No  & High & No  \\
\textbf{FDI}      & High & Yes & High & No  \\
\textbf{Proposed} & High & No  & High & No  \\
\hline
\end{tabular}
\vspace{-0.4cm}
\end{table}

Against this background, this paper studies cooperative OFDM-ISAC networks from the coupled perspectives of performance analysis and resource allocation. 
The main contributions are summarized as follows:
\begin{itemize}
\item \textbf{Fine-grained 2D orthogonal coordination:}
We establish a network-wide RE-level orthogonal coordination model for multiple Tx-BSs and adopt non-periodic 2D sensing-RE patterns to mitigate the effective bandwidth/CPI loss or periodic ambiguity replicas of conventional 1D orthogonal baselines.

\item \textbf{Performance analysis for SLF and PLF:}
We characterize the joint position/velocity sensing performance under two representative fusion architectures. For SLF, we derive the centralized CRB under the raw-observation model. For PLF, we develop a two-stage CRB-like metric by combining local delay/Doppler uncertainty characterization with first-order geometric error propagation. The PLF two-stage CRB framework covers both practical FFT-based local extraction and an oracle continuous-parameter ML benchmark, and clarifies that only the oracle ML-based PLF benchmark can asymptotically coincide with the SLF CRB under restrictive conditions.

\item \textbf{Time-frequency resource allocation design:}
Leveraging the above analysis and using the CRB as the transmitter-side sensing metric, we formulate a joint RE-selection and power-allocation problem under network-wide RE exclusivity, per-BS power budgets, a communication sum-rate constraint, and a sidelobe-amplitude constraint on the delay-Doppler ambiguity response. An efficient solution strategy is developed via Schur-complement reformulation and penalty-based alternating optimization.

\item \textbf{Numerical validation and design insights:}
Simulations validate the proposed SLF/PLF performance analysis and the CRB-based resource-allocation design. The proposed allocation design achieves consistent localization and velocity gains over representative orthogonal baselines. The results also show that the practical PLF loss mainly stems from finite-resolution local extraction and unweighted fusion, and that the SLF--PLF gap is strongly geometry dependent, being small in balanced deployments but much more pronounced in asymmetric regions.

\end{itemize}

\textit{Notation:} Boldface lower-case letters and bold upper-case letters denote vectors and matrices, respectively. The symbols $(\cdot)^{\ast}$, $(\cdot)^T$, $(\cdot)^H$, and $(\cdot)^{-1}$ denote conjugate, transpose, Hermitian transpose, and inverse, respectively. The sets $\mathbb{C}$, $\mathbb{R}$, and $\mathbb{Z}$ denote the complex, real, and integer domains. The $\ell_2$ norm is denoted by $\|\cdot\|_2$. The real part of a complex number is denoted by $\mathfrak{R}\{\cdot\}$, and $\mathbb{S}_N^{+}$ denotes the set of all $N$-dimensional complex positive semidefinite matrices. An $N\times N$ identity matrix is denoted by $\mathbf{I}_N$. The symbols $\mathbf{1}_N$ and $\mathbf{1}_{N\times M}$ denote the all-one vector and all-one matrix of compatible dimensions, respectively. The function $\mathrm{vec}\{\mathbf{A}\}$ vectorizes $\mathbf{A}$ column-wise, and $\mathrm{diag}\{\mathbf{a}\}$ forms a diagonal matrix from $\mathbf{a}$. The operators $\otimes$ and $\odot$ denote the Kronecker and Hadamard products, respectively.

\begin{figure}[t]
 \centering
   \vspace{-0.0 cm}
  \includegraphics[width=  3.4  in]{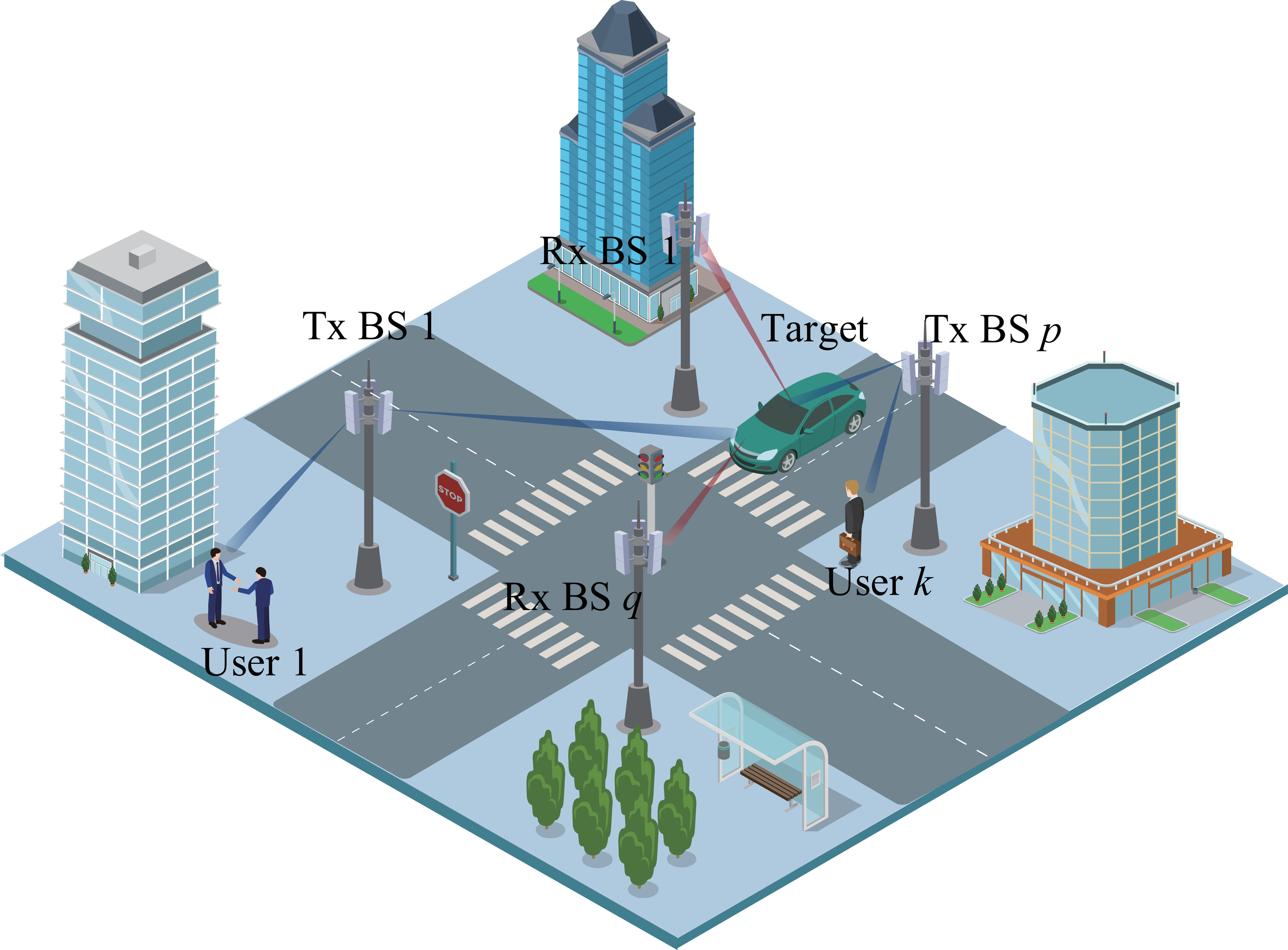}
  \caption{The considered cooperative OFDM-ISAC system.}\label{fig:system model}
    \vspace{-0.2 cm}
\end{figure}

\section{System Model and Problem Description}

We consider a cooperative OFDM-ISAC network illustrated in Fig.~\ref{fig:system model}, consisting of $P$ transmit BSs (Tx-BSs) and $Q$ receive BSs (Rx-BSs).
The $p$-th Tx-BS and the $q$-th Rx-BS are located at $\mathbf{b}_p=[x_p,\;y_p]^T$, $p=1,\ldots,P$, and $\mathbf{b}_q=[x_q,\;y_q]^T$, $q=1,\ldots,Q$, respectively.
The $P$ Tx-BSs cooperatively transmit OFDM waveforms to serve $K$ downlink communication users, while the $Q$ Rx-BSs simultaneously collect target echoes for multi-static sensing using the same OFDM waveforms.

\subsection{Transmit Signal Model with Orthogonal Time--Frequency Coordination}
Each Tx-BS adopts a common OFDM frame with $N$ subcarriers and $M$ OFDM symbols, where the $(n,m)$-th resource element (RE) corresponds to subcarrier $n$ in OFDM symbol $m$. We impose network-wide RE exclusivity across all Tx-BSs, i.e., each RE can be occupied by at most one Tx-BS to carry either a deterministic sensing pilot or a random communication information symbol. For sensing, orthogonal pilot-RE allocation across Tx-BSs enables the Rx-BSs to separate echoes from different Tx--target--Rx bistatic paths and estimate the associated delays and Dopplers, which are then fused for cooperative localization and velocity estimation. For downlink communication, we employ an OFDMA-type scheme that allocates mutually orthogonal RE subsets to different users, thereby avoiding inter-user interference.

To mathematically model the above coordination, let $\mathbf{A}_{p,0}\in\{0,1\}^{N\times M}$ denote the sensing-RE indicator matrix of Tx-BS $p$, where $a_{p,0,n,m}\triangleq \mathbf{A}_{p,0}(n,m)=1$ indicates that RE $(n,m)$ is used to transmit a sensing pilot. Similarly, let $\mathbf{A}_{p,k}\in\{0,1\}^{N\times M}$ denote the communication-RE indicator matrix for user $k$, where $a_{p,k,n,m}\triangleq \mathbf{A}_{p,k}(n,m)=1$ means that Tx-BS $p$ transmits a data symbol for user $k$ on RE $(n,m)$.
The network-wide orthogonal time--frequency coordination is enforced by the per-RE exclusivity constraint
\begin{equation}
\sum_{p=1}^P\Big(\mathbf{A}_{p,0}+\sum_{k=1}^K\mathbf{A}_{p,k}\Big)\preceq \mathbf{1}_{N\times M},
\end{equation}
which guarantees that each RE is occupied by at most one Tx-BS for either sensing or communication.
Let $\mathbf{S}_{p,0}\in\mathbb{C}^{N\times M}$ collect the sensing pilots and let $\mathbf{S}_{p,k}\in\mathbb{C}^{N\times M}$ collect the information symbols for user $k$.
The pilots and symbols are normalized such that $|\mathbf{S}_{p,0}(n,m)|^2=1$  and $\mathbb{E}\{|\mathbf{S}_{p,k}(n,m)|^2\}=1$. We further define the per-RE transmit power matrix of Tx-BS $p$ as $\mathbf{P}_p\in\mathbb{R}_+^{N\times M}$, where $p_{p,n,m}\triangleq \mathbf{P}_p(n,m)$ denotes the power allocated to RE $(n,m)$.

Given $\mathbf{A}_{p,0}$, $\mathbf{S}_{p,0}$, and $\mathbf{P}_p$, the sensing signal transmitted by the $p$-th Tx-BS can be expressed as
\begin{align}\label{eq:define Xp0}
\mathbf{X}_{p,0} \triangleq \mathbf{P}_p^{\odot \frac{1}{2}} \odot \mathbf{A}_{p,0}\odot \mathbf{S}_{p,0},
\end{align}
where $\mathbf{P}_p^{\odot \frac{1}{2}}$ denotes the element-wise square root of $\mathbf{P}_p$. Similarly, the communication signal for user $k$ is
\begin{align}
\mathbf{X}_{p,k} \triangleq \mathbf{P}_p^{\odot \frac{1}{2}} \odot \mathbf{A}_{p,k}\odot \mathbf{S}_{p,k}.
\end{align}
Then, the composite OFDM pilot/symbol matrix to be transmitted by Tx-BS $p$ can be expressed as
\begin{equation}\label{eq:Xp def}
\mathbf{X}_p = \mathbf{X}_{p,0} + \sum_{k=1}^K \mathbf{X}_{p,k}.
\end{equation}

Let $x_{p,n,m}\triangleq \mathbf{X}_p(n,m)$ denote the pilot/symbol transmitted by Tx-BS $p$ on RE $(n,m)$. By applying an $N$-point IFFT to $\{x_{p,n,m}\}_{n=0}^{N-1}$ and prepending a cyclic prefix (CP) of duration $T_{\mathrm{CP}}$ to each OFDM symbol, the frequency-domain symbols are mapped to a continuous-time complex baseband waveform. The resulting baseband signal can be written as
\begin{equation}\label{eq:ofdm_tx_time}
x_p(t)
\!=\! \frac{1}{\sqrt{N}}\!\! \sum_{m=0}^{M-1}\! \sum_{n=0}^{N-1} x_{p,n,m}\,
e^{\jmath 2\pi n\Delta f\,(t-mT_{\mathrm{s}}-T_{\mathrm{CP}})}
\,\tilde{g}(t-mT_{\mathrm{s}}),
\end{equation}
where $\Delta f$ is the subcarrier spacing, $T_{\mathrm{u}}=1/\Delta f$ is the useful OFDM symbol duration (excluding CP), $T_{\mathrm{s}}=T_{\mathrm{u}}+T_{\mathrm{CP}}$ is the total OFDM symbol duration, and $\tilde{g}(t)$ denotes a unit-amplitude rectangular pulse of duration $T_{\mathrm{s}}$.
The shift by $T_{\mathrm{CP}}$ ensures that for $t\in[mT_{\mathrm{s}},(m+1)T_{\mathrm{s}})$ the argument $(t-mT_{\mathrm{s}}-T_{\mathrm{CP}})\in[-T_{\mathrm{CP}},T_{\mathrm{u}})$, thereby capturing the cyclic-prefix extension of the useful OFDM symbol.

\subsection{Communication Signal Model and Performance Metric}

At user $k$, the received radio-frequency signal is down-converted to baseband and then processed by sampling, CP removal, and an $N$-point FFT for each OFDM symbol.
As a result, the received symbol on RE $(n,m)$ can be written as
\begin{equation}
y_{k,n,m}= \sum_{p=1}^{P} h_{p,k,n,m}\, x_{p,n,m} + z_{k,n,m},
\end{equation}
where $h_{p,k,n,m}$ denotes the downlink channel coefficient from Tx-BS $p$ to user $k$ on RE $(n,m)$, and
$z_{k,n,m}\sim \mathcal{CN}(0,\sigma_\text{comm}^2)$ is AWGN.

Under the proposed orthogonal coordination, each RE is assigned to at most one Tx-BS, and OFDMA ensures that each communication RE is used by at most one user.
Therefore, there is no inter-user (and inter-BS) interference on the same RE, and the instantaneous SNR contribution from Tx-BS $p$ to user $k$ on RE $(n,m)$ is
$|h_{p,k,n,m}|^2 p_{p,n,m} a_{p,k,n,m}/\sigma_\text{comm}^2$.
Accordingly, the achievable communication sum-rate averaged over all REs is
\begin{equation}
R_\mathrm{c}
= \frac{1}{MN}\sum_{p,k,n,m}
\log_2\left(1+\frac{|h_{p,k,n,m}|^2 p_{p,n,m} a_{p,k,n,m}}{\sigma_\text{comm}^2} \right).
\end{equation}

\subsection{Sensing Observation Signal Model}

Consider a point target located at Cartesian coordinate $\mathbf{u}=[x,\;y]^T$ and moving with a constant velocity $\mathbf{v}=[v_x,\;v_y]^T$ during one OFDM frame.
The complex baseband echo received at Rx-BS $q$ can be expressed as
\begin{equation}\label{eq:receive_signal_time_domain}
y_q(t)
= \sum_{p=1}^{P}\alpha_{p,q}\,
x_p(t-\tau_{p,q})\, e^{\jmath 2\pi f_{\mathrm{d},p,q}t}
+ z_q(t).
\end{equation}
Here $x_p(t)$ denotes the complex baseband signal transmitted by Tx-BS $p$. The scalar $\alpha_{p,q}$ is the complex reflection coefficient associated with the bistatic path from Tx-BS-$p$ to the target to Rx-BS-$q$. The parameters $\tau_{p,q}$ and $f_{\mathrm{d},p,q}$ are the corresponding bistatic propagation delay and Doppler shift, respectively. The complex AWGN is modeled as $z_q(t)\sim \mathcal{CN}(0,\sigma^2)$.
The delay and Doppler parameters in \eqref{eq:receive_signal_time_domain} are determined by the network geometry and the target state:
\begin{subequations}\label{eq:geometrical relationship}
\begin{align}
\tau_{p,q} &\triangleq \frac{1}{c_0}\big(\|\mathbf{u}-\mathbf{b}_p\|_2+\|\mathbf{u}-\mathbf{b}_q\|_2\big),\\
f_{\mathrm{d},p,q} &\triangleq \frac{1}{\lambda}
\left(\frac{\mathbf{v}^T(\mathbf{u}-\mathbf{b}_p)}{\|\mathbf{u}-\mathbf{b}_p\|_2}
+\frac{\mathbf{v}^T(\mathbf{u}-\mathbf{b}_q)}{\|\mathbf{u}-\mathbf{b}_q\|_2}\right),
\end{align}
\end{subequations}
where $\lambda=c_0/f_{\mathrm{c}}$ is the wavelength associated with the carrier frequency $f_\mathrm{c}$.
These expressions make explicit that the unknown target state $(\mathbf{u},\mathbf{v})$ induces a structured collection of per-link delay/Doppler parameters $\{\tau_{p,q}, f_{\mathrm{d},p,q}\}$ across all Tx-BS and Rx-BS pairs. Since these per-link quantities appear explicitly in the received echo model, it is convenient to introduce the intermediate parameterization
\begin{align}\label{eq:intermediate_parameter_eta_pq}
\bm{\eta}_{p,q} \triangleq [\tau_{p,q}, \;f_{\mathrm{d},p,q},\; \Re\{\alpha_{p,q}\},\;\Im\{\alpha_{p,q}\}]^T\in\mathbb{R}^4 .
\end{align}
Stacking $\bm{\eta}_{p,q}$ over all $(p,q)$ pairs yields the global intermediate vector
\begin{align}\label{eq:define_eta}
\bm{\eta}\triangleq [\bm{\tau}^T,\;\mathbf{f}_{\mathrm D}^T,\;\Re\{\bm{\alpha}\}^T,\;\Im\{\bm{\alpha}\}^T]^T\in \mathbb{R}^{4PQ},
\end{align}
where $\bm{\tau}\in\mathbb{R}^{PQ}$ and $\mathbf{f}_{\mathrm D}\in\mathbb{R}^{PQ}$ collect all $\tau_{p,q}$ and $f_{\mathrm{d},p,q}$, respectively.

Our ultimate target-related parameter vector is defined as
\begin{equation}\label{eq:global_parameter_xi}
\bm{\xi}\triangleq\big[\mathbf{u}^T,\;\mathbf{v}^T,\;\Re\{\bm{\alpha}\}^T,\;\Im\{\bm{\alpha}\}^T\big]^T \in \mathbb{R}^{4+2PQ},
\end{equation}
where $\bm{\alpha}\triangleq[\alpha_{1,1},\ldots,\alpha_{P,Q}]^T$ stacks the bistatic reflection coefficients for all $(p,q)$ pairs. Since $\tau_{p,q}$ and $f_{\mathrm{d},p,q}$ are deterministic functions of $(\mathbf{u},\mathbf{v})$ in \eqref{eq:geometrical relationship}, $\bm{\eta}$ can be viewed as a differentiable function of $\bm{\xi}$. This intermediate parameterization will be used later to derive the Fisher information from the echo model and then map it to $\bm{\xi}$ via a Jacobian transformation.

We now translate the continuous-time model in \eqref{eq:receive_signal_time_domain} to the OFDM time--frequency grid. Rx-BS $q$ samples $y_q(t)$ at $t=mT_{\mathrm{s}}+\ell/(N\Delta f)$, removes the cyclic prefix, and applies an $N$-point FFT on each OFDM symbol. Under the standard assumption $|f_{\mathrm{d},p,q}|\ll 0.1\Delta f$ for all $p,q$, inter-carrier interference is negligible. The resulting frequency-domain observation on RE $(n,m)$ can be written as
\begin{equation}
y_{q,n,m}
= \sum_{p=1}^{P}\alpha_{p,q}\,
e^{-\jmath 2\pi n \Delta f\,\tau_{p,q}}\,
e^{\jmath 2\pi m f_{\mathrm{d},p,q} T_{\mathrm{s}}}\,
x_{p,n,m}
+ z_{q,n,m},
\end{equation}
where $z_{q,n,m}\sim \mathcal{CN}(0,\sigma^2)$.
In the sensing pipeline, we only retain the resource elements allocated to sensing pilots, i.e., $\mathbf{X}_{p,0}$.
Stacking $\{y_{q,n,m}\}$ over subcarriers and OFDM symbols yields the compact matrix model
\begin{equation}\label{eq:received_signal_frequency}
\mathbf{Y}_q
= \sum_{p=1}^{P}
\Big(\alpha_{p,q}\,
\bm{\phi}(\tau_{p,q})\bm{\psi}^H(f_{\mathrm{d},p,q})
\odot \mathbf{X}_{p,0}\Big)
+ \mathbf{Z}_q,
\end{equation}
where $\mathbf{Y}_q\in\mathbb{C}^{N\times M}$ collects the received frequency-domain samples at Rx-BS $q$ and $\mathbf{Z}_q$ is the corresponding noise matrix. The delay steering vector and the Doppler steering vector are defined as
\begin{subequations}
\begin{align}
\bm{\phi}(\tau)&\triangleq
\big[1, e^{-\jmath2\pi \Delta f\tau}, \ldots,
e^{-\jmath2\pi (N-1)\Delta f\tau}\big]^T,  \\
\bm{\psi}(f_{\mathrm{d}})&\triangleq
\big[1, e^{-\jmath2\pi f_{\mathrm{d}}T_{\mathrm{s}}}, \ldots,
e^{-\jmath2\pi (M-1)f_{\mathrm{d}}T_{\mathrm{s}}}\big]^T.
\end{align}
\end{subequations}
This matrix form makes the dependence of the received data on the path parameters $\bm{\eta}_{p,q}$ explicit, and it also highlights how the transmitter-side sensing pattern $\mathbf{X}_{p,0}$ shapes the information available for estimation.

\subsection{Problem Statement and Design Scope}
Building on the above system model, this paper addresses two coupled tasks in cooperative OFDM-ISAC: receiver-side post-fusion performance analysis and transmitter-side time-frequency resource allocation. Under the network-wide RE-exclusivity constraint in~(1), sensing pilots and downlink data symbols compete for the same OFDM grid, and the system must jointly determine the sensing-RE pattern $\mathbf{A}_{p,0}$, the communication assignments $\{\mathbf{A}_{p,k}\}$, and the per-RE transmit powers $\mathbf{P}_p$  to enable accurate position/velocity sensing and reliable communication. Since the final sensing accuracy is determined after multi-BS fusion, Section III first characterizes the achievable position/velocity accuracy under SLF and PLF. Section IV then adopts the SLF CRB as the sensing design metric and develops joint RE selection and power allocation.

\vspace{-0.0 cm}
\section{Post-Fusion Performance Analysis under SLF and PLF}

\begin{figure}[t]
\centering
\vspace{-0.0cm}
\includegraphics[width=3.2in]{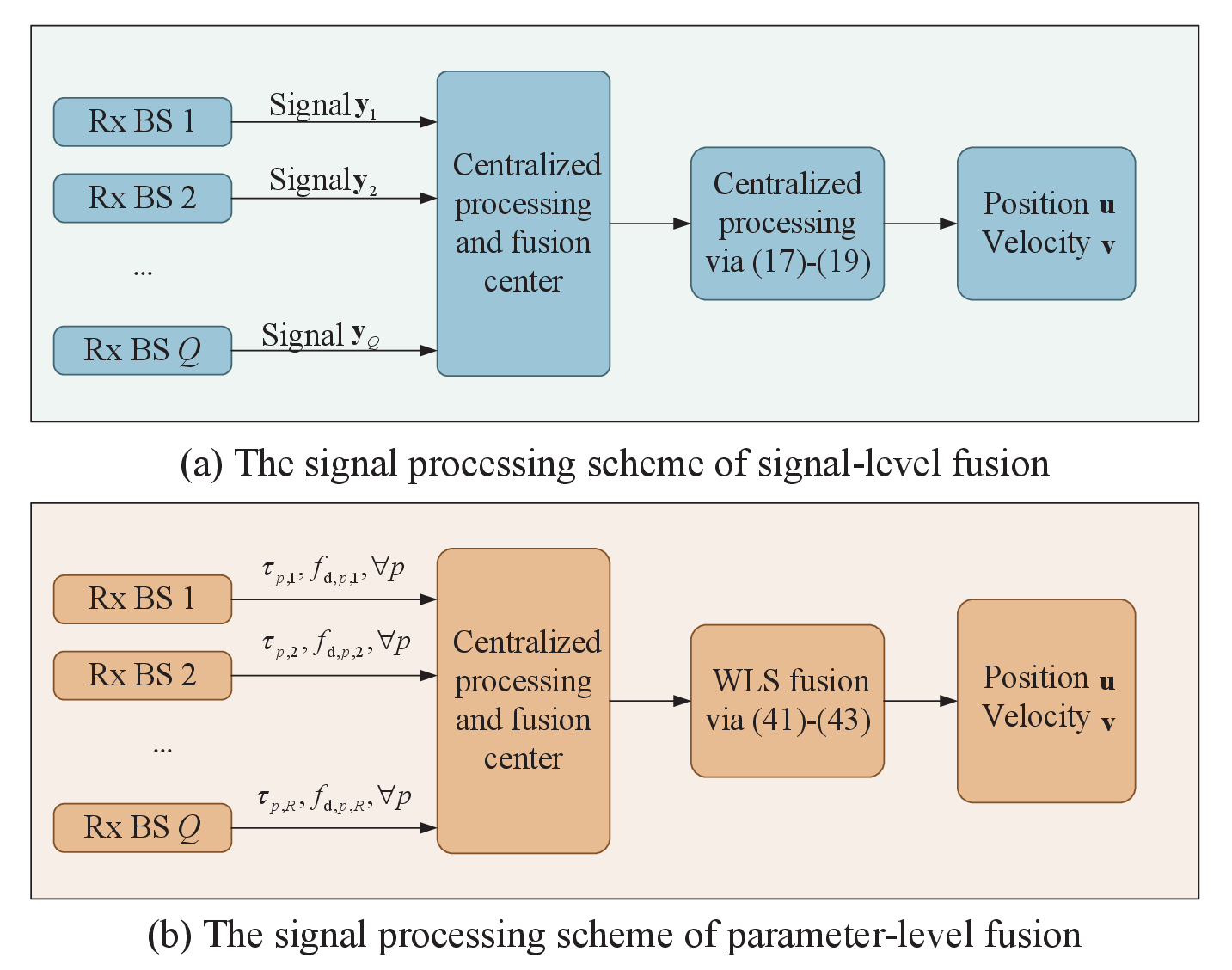}
\caption{The signal processing pipelines.}
\label{fig:fusion scheme}
\vspace{-0.2cm}
\end{figure}

This section establishes post-fusion sensing performance metrics for the two fusion architectures in Fig.~\ref{fig:fusion scheme}. For SLF, we derive the centralized CRB under the raw-observation model. For PLF, we characterize the local delay/Doppler report covariance and propagate it through geometric fusion to obtain a two-stage CRB-like metric. We then clarify the relation between the two metrics and identify the conditions under which PLF can asymptotically approach the SLF benchmark.

\subsection{The Estimation Methodologies and Cram\'er--Rao Bound Derivation for SLF}\label{subsec:CRB_signal}
Under SLF, the fusion center has access to the frequency-domain sensing observation matrices $\{\mathbf{Y}_q\}_{q=1}^Q$ defined in \eqref{eq:received_signal_frequency}. Each $\mathbf{Y}_q\in\mathbb{C}^{N\times M}$ is collected at the $q$-th Rx-BS over an OFDM frame with $N$ subcarriers and $M$ symbols and is backhauled to the fusion center for centralized processing. We define the vectorized observation at Rx-BS $q$ as $\mathbf{y}_q\triangleq \mathrm{vec}(\mathbf{Y}_q)$, and we stack all Rx BS observations as $\mathbf{y}\triangleq[\mathbf{y}_1^T,\dots,\mathbf{y}_Q^T]^T\in \mathbb{C}^{QMN}$. From \eqref{eq:received_signal_frequency}, the vectorized observation at the $q$-th Rx-BS can be written as
\begin{equation}\label{eq:y_r_vector_revised}
\mathbf{y}_{q}
= \sum_{p=1}^{P} \alpha_{p,q}\big(\bm{\psi}^{\ast}(f_{\mathrm{d},p,q})\otimes \bm{\phi}(\tau_{p,q})\big)\odot \mathbf{x}_{p,0} + \mathbf{z}_q,
\end{equation}
where $\mathbf{z}_q\sim \mathcal{CN}(\mathbf{0},\sigma^2\mathbf{I})$ denotes the vectorized AWGN.
The vector $\mathbf{x}_{p,0}\triangleq \mathrm{vec}(\mathbf{X}_{p,0})$ is the vectorized sensing waveform transmitted by Tx-BS $p$ over its allocated sensing REs.

Under the Gaussian noise assumption, the joint maximum-likelihood (ML) estimation of the target position $\mathbf{u}$, velocity $\mathbf{v}$, and complex reflection coefficient $\alpha_{p,q}$ can be formulated as a nonlinear least-squares problem:
\begin{equation}\label{eq:signal_level_ML}
\underset{\{\alpha_{p,q}\},\,\mathbf{u},\,\mathbf{v}}{\operatorname{arg\,min}}
\sum_{q=1}^{Q}\Big\|
\mathbf{y}_{q}-\sum_{p=1}^{P}\alpha_{p,q}
\big(\bm{\psi}^{\ast}(f_{\mathrm{d},p,q})\otimes \bm{\phi}(\tau_{p,q})\big)\odot \mathbf{x}_{p,0}
\Big\|_2^2 .
\end{equation}
With fixed $(\mathbf{u},\mathbf{v})$ and under network-wide RE exclusivity, the optimal $\alpha_{p,q}$ admits a closed-form solution by least squares:
\begin{equation}\label{eq:alpha_hat_signal}
\widehat{\alpha}_{p,q}=
\frac{
\big[\big(\bm{\psi}^{\ast}(f_{\mathrm{d},p,q})\otimes \bm{\phi}(\tau_{p,q})\big)\odot \mathbf{x}_{p,0}\big]^H\mathbf{y}_q
}{
\big\|\big(\bm{\psi}^{\ast}(f_{\mathrm{d},p,q})\otimes \bm{\phi}(\tau_{p,q})\big)\odot \mathbf{x}_{p,0}\big\|_2^2
}.
\end{equation}
Substituting \eqref{eq:alpha_hat_signal} into \eqref{eq:signal_level_ML} yields a concentrated cost function with respect to $(\mathbf{u},\mathbf{v})$:
\begin{equation}\label{eq:signal_level_concentrated}
\underset{\mathbf{u},\,\mathbf{v}}{\operatorname{arg\,min}}
\sum_{q=1}^{Q}\Big\|
\mathbf{y}_{q}\!-\!\sum_{p=1}^{P}\widehat{\alpha}_{p,q}
\big(\bm{\psi}^{\ast}(f_{\mathrm{d},p,q})\otimes \bm{\phi}(\tau_{p,q})\big)\odot \mathbf{x}_{p,0}
\Big\|_2^2.
\end{equation}

Since \eqref{eq:signal_level_concentrated} is generally non-convex and does not admit a closed-form solution, we adopt a two-stage strategy.
First, coarse estimates are obtained via a grid-based method (e.g., 2D-FFT peak search) followed by a simple least-squares fitting, which provides a reliable initialization.
Then, starting from the coarse solution, a high-precision estimate is obtained by directly minimizing \eqref{eq:signal_level_concentrated} using an off-the-shelf numerical optimizer; in particular, we employ a quasi-Newton update.
As SLF operates on raw echoes, it preserves the Fisher information and its estimator can asymptotically attain the CRB under standard regularity conditions.

To characterize the CRB for joint position and velocity estimation,
we require the FIM $\mathbf{F}_{\bm{\xi}}$ associated with $\bm{\xi}$ under the centralized observation model in \eqref{eq:y_r_vector_revised}.
Nevertheless, the likelihood implied by \eqref{eq:y_r_vector_revised} depends on $\bm{\xi}$ only through the per-link delay--Doppler parameters, which are explicitly collected in $\bm{\eta}$.
It is therefore convenient to first derive the FIM $\mathbf{F}_{\bm{\eta}}$ and then map it to $\bm{\xi}$ through the multistatic geometry.
Since $\bm{\eta}$ is a differentiable function of $\bm{\xi}$ through the multistatic geometry in Section~II, the chain rule of the FIM yields
\begin{equation}\label{eq:FIM_chain_revised}
\mathbf{F}_{\bm{\xi}}=\mathbf{J}^T\,\mathbf{F}_{\bm{\eta}}\,\mathbf{J},
\qquad
\mathbf{J}\triangleq \frac{\partial \bm{\eta}}{\partial \bm{\xi}^T}.
\end{equation}
The expressions of $\mathbf{F}_{\bm{\xi}}$ and $\mathbf{J}$ are provided in Appendix~\ref{appendix:slf_crb}.
Note that each element in $\mathbf{F}_{\bm{\xi}}$ is related to the masked power vectors
$\mathbf{p}_{p,0}=\mathrm{vec}(\mathbf{A}_{p,0}\odot \mathbf{P}_p)$.

Since only the position and velocity are of interest, we denote the target state of interest $\mathbf{s}$ and treat $\bm{\beta}$ as nuisance parameters, respectively defined by
\begin{equation}\label{eq:define_s_beta}
\begin{aligned}
\mathbf{s}&\triangleq [\mathbf{u}^T \; \mathbf{v}^T]^T =  [x,y,v_x,v_y]^T,\\
\bm{\beta}
&\triangleq
[\Re\{\bm{\alpha}^T\},\Im\{\bm{\alpha}^T\}]^T
\in\mathbb R^{2PQ},
\end{aligned}
\end{equation}
We partition $\mathbf{F}_{\bm{\xi}}$ as follows
\begin{equation}\label{eq:FIM_partition_revised}
\mathbf{F}_{\bm{\xi}} =
\left[\begin{array}{cc}
\mathbf{F}_{\mathbf{ss}} & \mathbf{F}_{\mathbf{s}\bm{\beta}}\\
\mathbf{F}_{\mathbf{s}\bm{\beta}}^T & \mathbf{F}_{\bm{\beta\beta}}
\end{array}\right],
\end{equation}
where $\mathbf{F}_{\mathbf{ss}}$ corresponds to target state $\mathbf{s}$ and $\mathbf{F}_{\bm{\beta\beta}}$ corresponds to $\bm{\beta}$.
By the Schur complement, the equivalent FIM for $\mathbf{s}$ is
\begin{equation}\label{eq:FIM_eff_revised}
\mathbf{F}_{\mathrm{eqv}}
=
\mathbf{F}_{\mathbf{ss}}
-
\mathbf{F}_{\mathbf{s}\bm{\beta}}
\mathbf{F}_{\bm{\beta}\bm{\beta}}^{-1}
\mathbf{F}_{\mathbf{s}\bm{\beta}}^T,
\end{equation}
and the CRB of joint position--velocity estimation is the inverse of equivalent FIM \cite{Kay_book-1993},
\begin{equation}
\mathrm{CRB}_\mathbf{s}^{\mathrm{SLF}}
\triangleq
\operatorname{tr}\!\left(\mathbf W \mathbf F_{\mathrm{eqv}}^{-1}\right),
\label{eq:WCRB}
\end{equation}
where $\mathbf{W}$ is the weighted matrix and is defined as
\begin{equation}
\mathbf{W}
=
\mathrm{diag}\!\left(
{\lambda_p}/{d_0^2},\;
{\lambda_p}/{d_0^2},\;
{\lambda_v}/{v_0^2},\;
{\lambda_v}/{v_0^2}
\right).
\end{equation}
Here, $d_0$ and $v_0$ denote the reference scales for position and velocity accuracy, respectively, while $\lambda_p$ and $\lambda_v$ are weighting coefficients that reflect the relative importance of localization and velocity estimation.
$\mathrm{CRB}_\mathbf{s}^{\mathrm{SLF}}$ denotes a weighted scalar CRB metric that characterizes the achievable estimation accuracy under SLF and will serve as the sensing objective in the subsequent resource allocation design.

\subsection{The Estimation Methodologies and Two-Stage Cram\'er--Rao Bound Derivation for PLF}

In PLF, each Rx-BS reports only local delay/Doppler parameters instead of raw observations.
The preserved information depends on the local extractor and the reported uncertainty statistics. We thus consider a unified PLF framework with a practical lossy FFT-based mode and an oracle asymptotically lossless ML-based mode.

\textit{Stage 1: Local delay/Doppler extraction.}
For each bistatic link $(p,q)$, define the local parameter of interest as
\begin{equation}
\bm{\zeta}_{p,q} \triangleq [\tau_{p,q},\, f_{\mathrm{d},p,q}]^T .
\label{eq:zeta_pq_def_new}
\end{equation}
The corresponding full local parameter vector is defined as \eqref{eq:intermediate_parameter_eta_pq}.
At the local extraction stage, two representative implementations are relevant.
First, in a practical FFT-based PLF receiver, the delay and Doppler are obtained from discretized spectral peak searching, e.g., by a 2D-FFT or related grid-based estimator, possibly followed by local refinement. In this case, the reported local statistic is inherently lossy because the raw signal is compressed onto a finite delay--Doppler grid before fusion. Denoting the reported estimate by $\hat{\bm{\zeta}}_{p,q}^{\mathrm{FFT}}$, we write
\begin{equation}
\hat{\bm{\zeta}}_{p,q}^{\mathrm{FFT}}
=
\bm{\zeta}_{p,q}
+
\bm{e}_{p,q}^{\mathrm{FFT}},
\label{eq:zeta_fft_model_new}
\end{equation}
with covariance
\begin{equation}
\bm{\Sigma}_{\zeta,p,q}^{\mathrm{FFT}}
\triangleq
\mathrm{Cov}\!\left(\bm{e}_{p,q}^{\mathrm{FFT}}\right).
\label{eq:sigma_fft_def_new}
\end{equation}
The error term $\bm{e}_{p,q}^{\mathrm{FFT}}$ contains not only the statistical estimation error caused by noise, but also the additional loss induced by grid discretization, peak-search mismatch, and finite FFT resolution. Therefore, the corresponding local report is generally information-lossy relative to the raw observation.

Second, in an oracle continuous-parameter ML-based PLF receiver, the local delay and Doppler are extracted by solving the continuous nonlinear least-squares problem
\begin{equation}\label{eq:parameter_level_ML}
\min_{\alpha_{p,q},\,\tau_{p,q},\,f_{\mathrm{d},p,q}}
\left\|
\mathbf{y}_{p,q}
-
\alpha_{p,q}
\big(\bm{\psi}^*(f_{\mathrm{d},p,q})\otimes\bm{\phi}(\tau_{p,q})\big)\odot \mathbf{x}_{p,0}
\right\|_2^2,
\end{equation}
where $\bm{y}_{p,q}$ stacks the received sensing samples associated with the $(p,q)$-th bistatic link. For fixed $(\tau_{p,q},f_{\mathrm{d},p,q})$, the optimal reflection coefficient is
\begin{equation}
\hat{\alpha}_{p,q}
=
\frac{
\left[
\big(\bm{\psi}^*(f_{\mathrm{d},p,q})\otimes\bm{\phi}(\tau_{p,q})\big)\odot \mathbf{x}_{p,0}
\right]^H\mathbf{y}_{p,q}
}{
\left\|
\big(\bm{\psi}^*(f_{\mathrm{d},p,q})\otimes\bm{\phi}(\tau_{p,q})\big)\odot \mathbf{x}_{p,0}
\right\|_2^2
}.
\label{eq:alpha_hat_param}
\end{equation}
Substituting \eqref{eq:alpha_hat_param} into \eqref{eq:parameter_level_ML} yields a two-dimensional continuous optimization in $(\tau_{p,q},f_{\mathrm{d},p,q})$, which can be solved by standard numerical optimizers.

Denote the resulting local ML estimate by
\begin{equation}
\hat{\bm{\zeta}}_{p,q}^{\mathrm{ML}}
=
\bm{\zeta}_{p,q}
+
\bm{e}_{p,q}^{\mathrm{ML}}.
\label{eq:zeta_ml_model_new}
\end{equation}
Under the standard regularity conditions and in the high-SNR/large-sample regime, the continuous-parameter ML estimator is asymptotically unbiased and efficient. Hence, its covariance approaches the equivalent local CRB:
\begin{equation}
\bm{\Sigma}_{\zeta,p,q}^{\mathrm{ML}}
\triangleq
\mathrm{Cov}\!\left(\bm{e}_{p,q}^{\mathrm{ML}}\right)
\approx
\left[\mathbf{F}^{-1}(\bm{\eta}_{p,q})\right]_{1:2,\,1:2}.
\label{eq:sigma_ml_cov_new}
\end{equation}
Let $\mathbf{F}(\bm{\eta}_{p,q})\in\mathbb{R}^{4\times 4}$ denote the local FIM of $\bm{\eta}_{p,q}$. As in Appendix A, it can be written as
\begin{equation}
[\mathbf{F}(\bm{\eta}_{p,q})]_{\ell,\ell'}
=
\frac{2}{\sigma^2}
\Re\!\left\{
\mathbf{p}_{p,0}^T \mathbf{g}^{(\ell,\ell')}_{p,q}
\right\},
\quad
\ell,\ell'\in\{1,2,3,4\},
\label{eq:local_fim_plf}
\end{equation}
where $\mathbf{p}_{p,0}=\mathrm{vec}(\mathbf{A}_{p,0}\odot \mathbf{P}_p)$, and $\mathbf{g}^{(\ell,\ell')}_{p,q}$ is defined by the same derivative vectors as in Appendix A. Partition $\mathbf{F}(\mathbf{\eta}_{p,q})$ as
\begin{equation}
\mathbf{F}(\bm{\eta}_{p,q})
=
\begin{bmatrix}
\mathbf{A}_{p,q} & \mathbf{B}_{p,q}\\
\mathbf{B}_{p,q}^T & \mathbf{C}_{p,q}
\end{bmatrix},
\label{eq:local_fim_partition_new}
\end{equation}
where $\mathbf{A}_{p,q}\in\mathbb{R}^{2\times 2}$ corresponds to $(\tau_{p,q},f_{\mathrm{d},p,q})$ and $\mathbf{C}_{p,q}\in\mathbb{R}^{2\times 2}$ corresponds to $(\Re\{\alpha_{p,q}\},\Im\{\alpha_{p,q}\})$. Then
\begin{equation}
\bm{\Sigma}_{\zeta,p,q}^{\mathrm{ML}}
\approx
(\mathbf{A}_{p,q}-\mathbf{B}_{p,q}\mathbf{C}_{p,q}^{-1}\mathbf{B}_{p,q}^T)^{-1}.
\label{eq:sigma_ml_equiv_crb_new}
\end{equation}

The above two local reporting modes lead to two different PLF operating points. The FFT-based report is practical but generally lossy, whereas the continuous-parameter ML-based report, when paired with exact covariance information, serves as an oracle asymptotically lossless benchmark for PLF. The latter does not mean that every compressed report is exactly lossless for finite samples; rather, it provides the benchmark case under which the local compression does not incur an additional first-order loss beyond the local CRB characterization.

\textit{Stage 2: Geometric fusion at the fusion center.}
At the fusion stage, stack all reported delay/Doppler pairs as
\begin{equation}
\hat{\bm{\zeta}}
=
\begin{bmatrix}
\hat{\bm{\tau}}\\
\hat{\mathbf{f}}_D
\end{bmatrix}
=
\left[
\hat{\tau}_{1,1},\ldots,\hat{\tau}_{P,Q},
\hat{f}_{d,1,1},\ldots,\hat{f}_{d,P,Q}
\right]^T
\in\mathbb{R}^{2PQ}.
\label{eq:zeta_stack_new}
\end{equation}
The geometry-based measurement function is
\begin{equation}
\bm{\zeta}(\mathbf{s})
\triangleq
[\bm{\tau}^T(\mathbf{u}),\, \mathbf{f}_\mathrm{D}^T(\mathbf{u},\mathbf{v})]^T,
\label{eq:zeta_s_new}
\end{equation}
where $\mathbf{s}=[\mathbf{u}^T,\mathbf{v}^T]^T$. Accordingly, the fusion-stage measurement model is
\begin{equation}
\hat{\bm{\zeta}}
=
\bm{\zeta}(\mathbf{s})
+
\bm{\epsilon},
\label{eq:fusion_measurement_model_new}
\end{equation}
where $\bm{\epsilon}$ denotes the stacked local extraction error. Since different bistatic links are extracted from orthogonally allocated sensing REs, we neglect cross-link correlations and write
\begin{equation}
\bm{\Sigma}_{\zeta}
=
\mathrm{Cov}(\bm{\epsilon})
=
\mathrm{blkdiag}\!\left(
\bm{\Sigma}_{\zeta,1,1},
\ldots,
\bm{\Sigma}_{\zeta,P,Q}
\right).
\label{eq:blockdiag_sigma_zeta_new}
\end{equation}
Here, $\bm{\Sigma}_{\zeta,p,q}$ can represent either the practical FFT-based covariance $\bm{\Sigma}_{\zeta,p,q}^{\mathrm{FFT}}$ or the oracle ML-based covariance $\bm{\Sigma}_{\zeta,p,q}^{\mathrm{ML}}$, depending on the adopted local extractor.

Under the first-order linearization of \eqref{eq:fusion_measurement_model_new} around the true state,
\begin{equation}
\hat{\bm{\zeta}}
\approx
\bm{\zeta}(\mathbf{s}_0)
+
\mathbf{J}_\mathbf{s}(\mathbf{s}-\mathbf{s}_0)
+
\bm{\epsilon},
\label{eq:first_order_linearization_new}
\end{equation}
where $\mathbf{J}_\mathbf{s}$ denotes the Jacobian of $\bm{\zeta}(\bm{s})$ with respect to $\bm{s}$. To account for heterogeneous link qualities, we adopt weighted least-squares (WLS) fusion:
\begin{equation}
\hat{\mathbf{s}}_{\mathrm{WLS}}
=
\arg\min_{\mathbf{s}}
\left(
\hat{\bm{\zeta}}-\bm{\zeta}(\mathbf{s})
\right)^T
\mathbf{W}_{\mathrm{WLS}}
\left(
\hat{\bm{\zeta}}-\bm{\zeta}(\mathbf{s})
\right),
\label{eq:wls_estimator_new}
\end{equation}
with $\mathbf{W}_{\mathrm{WLS}}\succ \bm{0}$. The resulting first-order estimation error is
\begin{equation}
\hat{\mathbf{s}}_{\mathrm{WLS}}-\mathbf{s}
\approx
(\mathbf{J}_\mathbf{s}^T\mathbf{W}_{\mathrm{WLS}}\mathbf{J}_\mathbf{s})^{-1}
\mathbf{J}_\mathbf{s}^T\mathbf{W}_{\mathrm{WLS}}\bm{\epsilon},
\label{eq:wls_error_new}
\end{equation}
and its covariance is
\begin{equation}
\begin{aligned}
\bm{\Sigma}_\mathbf{s}
&\triangleq
\mathbb{E}
\!\left[
(\hat{\mathbf{s}}_{\mathrm{WLS}}-\mathbf{s})
(\hat{\mathbf{s}}_{\mathrm{WLS}}-\mathbf{s})^T
\right]\\
&\approx
(\mathbf{J}_\mathbf{s}^T\mathbf{W}_{\mathrm{WLS}}\mathbf{J}_\mathbf{s})^{-1}
\mathbf{J}_\mathbf{s}^T\mathbf{W}_{\mathrm{WLS}}
\bm{\Sigma}_{\zeta}
\mathbf{W}_{\mathrm{WLS}}\mathbf{J}_\mathbf{s}
(\mathbf{J}_\mathbf{s}^T\mathbf{W}_{\mathrm{WLS}}\mathbf{J}_\mathbf{s})^{-1}.
\label{eq:sigma_s_general_new}
\end{aligned}
\end{equation}
With the statistically consistent choice
$\mathbf{W}_{\mathrm{WLS}}=\bm{\Sigma}_{\zeta}^{-1}$,
we obtain
$
\bm{\Sigma}_\mathbf{s}
\approx
(\mathbf{J}_\mathbf{s}^T\bm{\Sigma}_{\zeta}^{-1}\mathbf{J}_\mathbf{s})^{-1}$.

This motivates the following unified architecture-dependent PLF performance metric:
\begin{equation}
\mathrm{TS\mbox{-}CRB}_\mathbf{s}(\bm{\Sigma}_{\zeta})
\triangleq
\mathrm{tr}\!\left(
\mathbf{W}\,
(\mathbf{J}_\mathbf{s}^T\bm{\Sigma}_{\zeta}^{-1}\mathbf{J}_\mathbf{s})^{-1}
\right),
\label{eq:general_tscrb_new}
\end{equation}
where $\mathbf{W}$ is the same weighting matrix used in the SLF case.
If $\bm{\Sigma}_{\zeta}$ is formed from the practical FFT-based local extraction covariance in \eqref{eq:sigma_fft_def_new}, then \eqref{eq:general_tscrb_new} yields the practical TS-CRB of a lossy PLF architecture. This metric reflects both local noise-induced estimation error and the additional information loss caused by grid-based compression.

If $\bm{\Sigma}_{\zeta}$ is formed from the equivalent local CRB blocks in \eqref{eq:sigma_ml_equiv_crb_new}, i.e.,
\begin{equation}
\bm{\Sigma}_{\zeta}^{\mathrm{ML}}
=
\mathrm{blkdiag}\!\left(
\bm{\Sigma}_{\zeta,1,1}^{\mathrm{ML}},
\ldots,
\bm{\Sigma}_{\zeta,P,Q}^{\mathrm{ML}}
\right),
\label{eq:sigma_zeta_ml_stack_new}
\end{equation}
then \eqref{eq:general_tscrb_new} becomes the oracle ML-based TS-CRB:
\begin{equation}
\mathrm{TS\mbox{-}CRB}_{\mathbf{s}}^{\mathrm{ML}}
\triangleq
\mathrm{tr}\!\left(
\mathbf{W}\,
(\mathbf{J}_{\mathbf{s}}^T(\bm{\Sigma}_{\zeta}^{\mathrm{ML}})^{-1}\mathbf{J}_{\mathbf{s}})^{-1}
\right).
\label{eq:tscrb_ml_oracle_new}
\end{equation}
This oracle metric is the appropriate benchmark for assessing whether PLF can asymptotically approach the SLF CRB. By contrast, the practical FFT-based TS-CRB should in general be expected to be looser, since the reported local statistic is information-lossy before the fusion stage.

Therefore, throughout the sequel, unless otherwise stated, the term \emph{TS-CRB} refers to the general PLF metric in \eqref{eq:general_tscrb_new}, while \emph{oracle ML-based TS-CRB} refers specifically to \eqref{eq:tscrb_ml_oracle_new}. Only the latter can asymptotically coincide with the SLF CRB under the restrictive lossless conditions clarified in Sec.~III-C.

\subsection{Performance Gap Analysis between  SLF and PLF}

We next clarify the relationship between the SLF CRB and the PLF TS-CRB.
The key point is that PLF is not unique: its achievable accuracy depends on whether the reported local delay/Doppler statistics are asymptotically lossless or inherently lossy.
Accordingly, we distinguish the following three objects:
i) the exact CRB under SLF, which is based on the raw observation;
ii) the oracle ML-based TS-CRB under PLF, where the local continuous-parameter extractor is asymptotically efficient and the fusion center uses the exact covariance;
iii) the practical FFT-based TS-CRB under PLF, where the reported local statistic is lossy due to grid discretization and finite FFT resolution.

Since the reported parameter vector $\hat{\bm{\zeta}}$ is obtained from the raw observation $\mathbf{y}$ through local estimation and compression, the variables form the Markov chain
\begin{equation}\label{eq:markov_chain_s_y_zeta_revised}
\mathbf{s}
\rightarrow
\{\mathbf{y}_q\}_{q=1}^{Q}
\rightarrow
\hat{\bm{\zeta}}.
\end{equation}
By the information-processing principle for estimation, post-processing cannot increase the information about $\mathbf{s}$.
Therefore, the Fisher information carried by the compressed parameter-domain observation cannot exceed that contained in the raw signal-domain observation.

Recall from Sec.~III-A that, after eliminating the nuisance parameter $\bm{\beta}$ via the Schur complement, the equivalent FIM of SLF can be written as
\begin{equation}\label{eq:Feq_SLF_markov_revised}
\mathbf{F}_{\mathrm{eqv}}^{\mathrm{SLF}}
=
\mathbf{J}_{\mathbf{s}}^{T}
\mathbf{G}_{\bm{\zeta}}
\mathbf{J}_{\mathbf{s}},
\end{equation}
where
$\mathbf{G}_{\bm{\zeta}}
\triangleq
\mathbf{F}_{\bm{\zeta}\bm{\zeta}}
-
\mathbf{F}_{\bm{\zeta}\bm{\beta}}
\mathbf{F}_{\bm{\beta}\bm{\beta}}^{-1}
\mathbf{F}_{\bm{\zeta}\bm{\beta}}^{T}$.

For a generic covariance-aware PLF receiver, if the reported local parameter vector has covariance $\bm{\Sigma}_{\bm{\zeta}}$ and the fusion center uses the exact covariance-aware WLS rule, then the corresponding parameter-domain FIM is
\begin{equation}\label{eq:F_PLF_generic_revised}
\mathbf{F}_{\mathbf{s}}^{\mathrm{PLF}}
=
\mathbf{J}_{\mathbf{s}}^{T}
\bm{\Sigma}_{\bm{\zeta}}^{-1}
\mathbf{J}_{\mathbf{s}}.
\end{equation}
By \eqref{eq:markov_chain_s_y_zeta_revised}, one has the matrix inequality
\begin{equation}\label{eq:FIM_ineq_main_revised}
\mathbf{F}_{\mathrm{eqv}}^{\mathrm{SLF}}
\succeq
\mathbf{F}_{\mathbf{s}}^{\mathrm{PLF}}.
\end{equation}
Since both matrices are positive definite, \eqref{eq:FIM_ineq_main_revised} implies
\begin{equation}\label{eq:CRB_matrix_ineq_revised}
\left(
\mathbf{F}_{\mathrm{eqv}}^{\mathrm{SLF}}
\right)^{-1}
\preceq
\left(
\mathbf{F}_{\mathbf{s}}^{\mathrm{PLF}}
\right)^{-1}.
\end{equation}
Using the same weighting matrix $\mathbf{W}\succeq\mathbf{0}$ as in the SLF case, the corresponding weighted bounds satisfy
\begin{equation}\label{eq:weighted_bound_ineq_revised}
\mathrm{CRB}_{\mathbf{s}}^{\mathrm{SLF}}
=
\operatorname{tr}
\!\left(
\mathbf{W}
\left(
\mathbf{F}_{\mathrm{eqv}}^{\mathrm{SLF}}
\right)^{-1}
\right)
\le
\operatorname{tr}
\!\left(
\mathbf{W}
\left(
\mathbf{F}_{\mathbf{s}}^{\mathrm{PLF}}
\right)^{-1}
\right).
\end{equation}

Therefore, for any covariance-aware PLF architecture described by $\bm{\Sigma}_{\bm{\zeta}}$, the corresponding first-order PLF metric can be written as
\begin{equation}\label{eq:TSCRB_general_revised}
\mathrm{TS\mbox{-}CRB}_{\mathbf{s}}(\bm{\Sigma}_{\bm{\zeta}})
\triangleq
\operatorname{tr}
\!\left(
\mathbf{W}
\left(
\mathbf{J}_{\mathbf{s}}^{T}
\bm{\Sigma}_{\bm{\zeta}}^{-1}
\mathbf{J}_{\mathbf{s}}
\right)^{-1}
\right),
\end{equation}
and \eqref{eq:weighted_bound_ineq_revised} becomes
\begin{equation}\label{eq:SLF_leq_general_PLF_revised}
\mathrm{CRB}_{\mathbf{s}}^{\mathrm{SLF}}
\le
\mathrm{TS\mbox{-}CRB}_{\mathbf{s}}(\bm{\Sigma}_{\bm{\zeta}}).
\end{equation}

We now specialize \eqref{eq:SLF_leq_general_PLF_revised} to the two representative local reporting modes in Sec.~III-B.

\textit{1) Oracle ML-based PLF:}
when the local delay/Doppler extractor is a continuous-parameter ML estimator operating in the asymptotically efficient regime, and the reported covariance is given by the equivalent local CRB blocks, we denote the stacked covariance by $\bm{\Sigma}_{\bm{\zeta}}^{\mathrm{ML}}$.
The corresponding oracle PLF benchmark is
\begin{equation}\label{eq:TSCRB_ML_revised}
\begin{aligned}
\mathrm{TS\mbox{-}CRB}_{\mathbf{s}}^{\mathrm{ML}}
&\triangleq
\mathrm{TS\mbox{-}CRB}_{\mathbf{s}}
\!\left(
\bm{\Sigma}_{\bm{\zeta}}^{\mathrm{ML}}
\right)\\
&=
\operatorname{tr}
\!\left(
\mathbf{W}
\left(
\mathbf{J}_{\mathbf{s}}^{T}
\left(
\bm{\Sigma}_{\bm{\zeta}}^{\mathrm{ML}}
\right)^{-1}
\mathbf{J}_{\mathbf{s}}
\right)^{-1}
\right).
\end{aligned}
\end{equation}
In this case,
\begin{equation}\label{eq:SLF_leq_ML_revised}
\mathrm{CRB}_{\mathbf{s}}^{\mathrm{SLF}}
\le
\mathrm{TS\mbox{-}CRB}_{\mathbf{s}}^{\mathrm{ML}}.
\end{equation}

The inequality in \eqref{eq:SLF_leq_ML_revised} becomes an equality only under restrictive conditions:
i) different bistatic links are statistically separable under the adopted orthogonal RE allocation;
ii) after nuisance elimination, the reported local statistic is asymptotically lossless for $\mathbf{s}$, i.e., it is an asymptotically sufficient statistic of the raw observation with respect to the target state;
iii) the local continuous-parameter ML estimator is asymptotically unbiased and efficient, such that its covariance achieves the equivalent local CRB;
iv) the fusion center uses the exact covariance-aware WLS rule, and the measurement-domain model is exact or asymptotically exact.

Under these conditions, one has $\bm{\Sigma}_{\bm{\zeta}}^{\mathrm{ML}}
=\mathbf{G}_{\bm{\zeta}}^{-1}$,
which yields
\begin{equation}
\mathbf{F}_{\mathbf{s}}^{\mathrm{PLF,ML}}
=
\mathbf{J}_{\mathbf{s}}^{T}
\mathbf{G}_{\bm{\zeta}}
\mathbf{J}_{\mathbf{s}}
=
\mathbf{F}_{\mathrm{eqv}}^{\mathrm{SLF}}.
\end{equation}
Hence, only the oracle ML-based TS-CRB can asymptotically coincide with the SLF CRB, i.e. $\mathrm{CRB}_{\mathbf{s}}^{\mathrm{SLF}}
=
\mathrm{TS\mbox{-}CRB}_{\mathbf{s}}^{\mathrm{ML}}$.

\textit{2) Practical FFT-based PLF:}
when the local delay/Doppler pair is extracted from grid-based spectral peak searching, such as a 2D-FFT estimator with finite FFT resolution, the reported local statistic is compressed onto a discrete delay--Doppler grid before fusion.
Denote the corresponding stacked covariance by $\bm{\Sigma}_{\bm{\zeta}}^{\mathrm{FFT}}$.
Then the practical FFT-based PLF metric is
\begin{equation}\label{eq:TSCRB_FFT_revised}
\begin{aligned}
\mathrm{TS\mbox{-}CRB}_{\mathbf{s}}^{\mathrm{FFT}}
&\triangleq
\mathrm{TS\mbox{-}CRB}_{\mathbf{s}}
\!\left(
\bm{\Sigma}_{\bm{\zeta}}^{\mathrm{FFT}}
\right)\\
&=
\operatorname{tr}
\!\left(
\mathbf{W}
\left(
\mathbf{J}_{\mathbf{s}}^{T}
\left(
\bm{\Sigma}_{\bm{\zeta}}^{\mathrm{FFT}}
\right)^{-1}
\mathbf{J}_{\mathbf{s}}
\right)^{-1}
\right).
\end{aligned}
\end{equation}
Because $\bm{\Sigma}_{\bm{\zeta}}^{\mathrm{FFT}}$ contains not only noise-induced estimation error but also additional discretization and peak-search loss, the corresponding local report is generally information-lossy relative to the raw observation.
Therefore, even when covariance-aware WLS is used at the fusion center, the practical FFT-based TS-CRB should in general be interpreted as a looser PLF bound than the oracle ML-based one, rather than as an asymptotically equivalent surrogate of the SLF CRB.

In summary, the exact SLF CRB provides the fundamental benchmark based on raw observations; the oracle ML-based TS-CRB is the appropriate asymptotically lossless PLF benchmark and can coincide with the SLF CRB only under restrictive conditions; and the practical FFT-based TS-CRB corresponds to a lossy PLF architecture and generally remains looser due to local grid-induced information loss.

\section{CRB-Oriented Time-Frequency Resource Allocation}
Motivated by the analysis in Section III, we optimize the transmitter-side time-frequency allocation using the SLF CRB as the sensing metric, since it is the information-preserving benchmark that depends explicitly on the sensing RE masks and power variables. The PLF metrics derived in Section III are used to quantify the performance loss caused by low-dimensional fusion and to interpret the architecture-dependent simulation results. Since the CRB captures only local estimation accuracy, we additionally impose a sidelobe-amplitude constraint to control off-peak ambiguity responses.

\subsection{Ambiguity Sidelobe Metric and Sidelobe-Amplitude Constraint}
CRB captures only local estimation accuracy around the true delay-Doppler pair and does not control off-peak ambiguity responses. To avoid allocations that are locally sharp but globally ambiguity-prone, we additionally impose a sidelobe-amplitude constraint on a discretized delay-Doppler lattice.

Specifically, for the $p$-th Tx-BS, consider the ambiguity response evaluated on integer-lattice offsets $(l,\nu)\in\mathbb{Z}^2$, where $l$ and $\nu$ denote the delay-bin and Doppler-bin indices, respectively.
We define the sidelobe-amplitude sample as
\begin{subequations}\label{eq:PSL_def}
\begin{align}
\Gamma_{p}(l,\nu)
&\triangleq \frac{1}{MN}\!\sum_{m=0}^{M-1}\sum_{n=0}^{N-1}
a_{p,0,n,m}\,p_{p,n,m}\,
e^{-\jmath 2\pi l\frac{n}{N}}
e^{\jmath 2\pi \nu\frac{m}{M}} \label{eq:PSL_scalar}\\
&=\frac{1}{MN}\left(\widetilde{\bm{\psi}}_\nu^{H}\otimes \widetilde{\bm{\phi}}_{l}\right)^{T}
\left(\mathbf{a}_{p,0}\odot \mathbf{p}_{p}\right), \label{eq:PSL_vec}
\end{align}
\end{subequations}
where $\widetilde{\bm{\phi}}_{l}\triangleq \bm{\phi}\big(\frac{l}{N\Delta f}\big)\in\mathbb{C}^{N}$ and
$\widetilde{\bm{\psi}}_{\nu}\triangleq \bm{\psi}\big(\frac{\nu}{MT_{\mathrm{s}}}\big)\in\mathbb{C}^{M}$
are the sampled steering vectors in the delay and Doppler domains, respectively.
Moreover, $\mathbf{a}_{p,0}\triangleq \mathrm{vec}(\mathbf{A}_{p,0})$ denotes the vectorized sensing-RE selection indicator, and
$\mathbf{p}_{p}\triangleq \mathrm{vec}(\mathbf{P}_{p})$ collects the corresponding per-RE transmit powers.
Equation \eqref{eq:PSL_vec} follows from vectorizing the two-dimensional DFT kernel across the time--frequency grid.

Based on \eqref{eq:PSL_def}, we enforce a sidelobe-amplitude constraint on a prescribed lattice set
$\mathcal{S}\triangleq\{(l,\nu):|l|\leq l_\mathrm{max}, |\nu|\leq\nu_{\mathrm{max}}, (l,\nu)\neq (0,0)\}$, which defines a finite window of delay-Doppler bins:
\begin{equation}\label{eq:PSL_constraint}
\max_{(l,\nu)\in\mathcal{S}} \big|\Gamma_{p}(l,\nu)\big|\leq \beta_0, \quad \forall p,
\end{equation}
where $\beta_0\in(0,1)$ is the threshold on the absolute sidelobe amplitude. For convenience, we also define its dB-form counterpart as
$\beta_{0,\mathrm{dB}} \triangleq 20\log_{10}(\beta_0)$.
This constraint complements the CRB-based objective by suppressing off-peak ambiguity responses, which enhances robustness against false alarms in adverse conditions.

\subsection{Problem Formulation and Algorithm}

In this section, we aim to minimize the CRB of position and velocity by jointly designing the time--frequency resource selection $\mathbf{a}_{p,0}$, $\mathbf{a}_{p,k}$, $\forall k$, and the power allocation $\mathbf{p}_{p}$, while satisfying the total transmit power budget, the communication sum-rate requirement, and the delay--Doppler sidelobe-amplitude constraint.
The resulting optimization problem is formulated as
\begin{subequations}\label{eq:original problem}
\begin{align}
\min_{\{\mathbf{a}_{p,0}, \mathbf{a}_{p,k}\}, \mathbf{p}_{p}, \forall p} & \ \mathrm{CRB}_{\mathbf{s}}^{\mathrm{SLF}} \label{eq:original problem obj}\\
\mathrm{s.t.}\ \ & \mathbf{p}_{p}^T \mathbf{1}_{MN}\leq P_{\mathrm{tot},p},\ \mathbf{p}_{p} \geq \mathbf{0},  \quad \forall p,  \label{eq:original problem c1}\\
&\ R_\mathrm{c} \geq \eta_0, \label{eq:original problem c3} \\
&\ \max_{(l,\nu)\in\mathcal{S}} \big|\Gamma_{p}(l,\nu)\big|\leq \beta_0, \quad \forall p, \label{eq:original problem c4}\\&\ \sum_{p=1}^{P}\big(\mathbf{a}_{p,0}+\sum_{k=1}^{K}\mathbf{a}_{p,k}\big)\preceq \mathbf{1}_{MN},  \label{eq:original problem c2}\\
&\  \mathbf{a}_{p,0},\mathbf{a}_{p,k} \in \{0,1\}^{MN},\ \forall p,k, \label{eq:original problem c5}
\end{align}
\end{subequations}
where $ P_{\mathrm{tot},p}$ denotes the total transmit power budget of BS $p$, $\eta_0$ is the communication sum-rate threshold, and $\beta_0$ is the delay--Doppler sidelobe amplitude threshold.
It is evident that problem~\eqref{eq:original problem} is highly non-convex due to the inverse-form CRB objective~\eqref{eq:original problem obj}, and the binary selection constraints~\eqref{eq:original problem c5}.
To tackle this difficulty, we subsequently develop an alternating optimization framework that updates the resource selection and power allocation in a coordinated manner.

\begin{algorithm}[!t]
\begin{small}
\caption{SLF-Based Resource Selection and Power Allocation Algorithm}
\label{alg1}
    \begin{algorithmic}[1]
    \REQUIRE  $P_{\mathrm{tot},p}$, $\beta_0$, $\eta_0$, $\mathbf{b}_p, \forall p$, $\mathbf{b}_q, \forall q$, $\mathbf{u}$, $\mathbf{v}$, $\nu_\mathrm{max}$, $l_\mathrm{max}$, $\sigma^2_z$, $h_{p,k,n,m}, \forall p, \forall k, \forall n, \forall m$, and $\sigma^2_\mathrm{comm}$.
    \ENSURE $\mathbf{p}_{p}^\star$, $\mathbf{a}_{p,0}^\star$ and $\mathbf{a}_{p,k}^\star$.
        \STATE {Initialize $\mathbf{p}_{p}$, $\mathbf{a}_{p,0}$, $\mathbf{a}_{p,k}$ and set $\rho_1$.}
        \WHILE {no convergence }
            \STATE{Update $\mathbf{p}_{p}$ and $\mathbf{L}$ by solving (\ref{eq:update p}).}
            \STATE{Update $\mathbf{a}_{p,0}$,$\mathbf{a}_{p,k}$ and $\mathbf{L}$ using (\ref{eq:update u}).}
        \ENDWHILE
        \STATE Perform final binary recovery on $\mathbf{a}_{p,0}$ and $\mathbf{a}_{p,k}$ for each RE.
        \STATE Update $\mathbf{p}_{p}$ and $\mathbf{L}$ by solving (\ref{eq:update p}).
        \STATE{Return $\mathbf{p}_{p}^\star =\mathbf{p}_{p}$, $\mathbf{a}_{p,0}^\star =\mathbf{a}_{p,0}$ and $\mathbf{a}_{p,k}^\star =\mathbf{a}_{p,k}$.}
    \end{algorithmic}
    \end{small}
    \vspace{-0.1cm}
\end{algorithm}

\subsubsection{Schur Complement Transformation}

To handle the inverse-form CRB objective, we introduce an auxiliary matrix variable $\mathbf{L}\in \mathbb{S}_{4}^{+}$ such that $\mathbf{L}\preceq  \mathbf{F}_{\mathbf{ss}}-\mathbf{F}_{\mathbf{s}\bm{\beta}}\mathbf{F}_{\bm{\beta}\bm{\beta}}^{-1}\mathbf{F}_{\mathbf{s}\bm{\beta}}^T$, and problem \eqref{eq:original problem} is transformed via Schur complement \cite{Schur_Complement} to
\begin{subequations}\label{eq:Schur transformation}
\begin{align}
\min_{\{\mathbf{a}_{p,0}, \mathbf{a}_{p,k}\}, \mathbf{p}_{p}, \forall p, \, \mathbf{L}} & \quad  \mathrm{Tr}\{\mathbf{W}\mathbf{L}^{-1}\}\label{eq:Schur transformation obj} \\
\mathrm{s.t.} & \quad  \left[ \begin{array}{ll}
\mathbf{F}_{\mathbf{ss}}-\mathbf{L} & \mathbf{F}_{\mathbf{s}\bm{\beta}} \\
\mathbf{F}_{\mathbf{s}\bm{\beta}}^T & \mathbf{F}_{\bm{\beta}\bm{\beta}}
\end{array}
\right]\succeq \mathbf{0}\label{eq:Schur transformation c1},\\
& \mathbf{p}_{p}^T \mathbf{1}_{MN}\leq P_{\mathrm{tot},p},\ \mathbf{p}_{p} \geq \mathbf{0}, \quad  \forall p, \\
&\ \sum_{p=1}^{P}\big(\mathbf{a}_{p,0}+\sum_{k=1}^{K}\mathbf{a}_{p,k}\big)\preceq \mathbf{1}_{MN}\label{eq:Schur transformation c3},\\
&\ R_\mathrm{c} \geq \eta_0, \\
&\ \max_{(l,\nu)\in\mathcal{S}} \big|\Gamma_{p}(l,\nu)\big|\leq \beta_0, \quad \forall p, \\
&\  \mathbf{a}_{p,0},\mathbf{a}_{p,k} \in \{0,1\}^{MN},\ \forall p,k \label{eq:Schur transformation c6}.
\end{align}
\end{subequations}

\subsubsection{Binary Integer Constraint Relaxation}
To handle the binary selection variables \eqref{eq:Schur transformation c6}, we relax it to the box constraint $\mathbf{0}_{MN}\preceq \mathbf{a}_{p,k}\preceq \mathbf{1}_{MN}, \forall p, k$, while preserving the network-wide RE-exclusivity constraint in its original linear form.
To further promote binary solutions, we add the concave penalty $g(\mathbf{a}_{p,k})\triangleq \mathbf{a}^T_{p,k}(\mathbf{1}_{MN}-\mathbf{a}_{p,k})$, which vanishes at binary points and is strictly positive for fractional entries.
Consequently, the problem can be transformed into the following form:
\begin{subequations}\label{eq:binary integer constraint relaxation}
\begin{align}
\min_{\{\mathbf{a}_{p,0}, \mathbf{a}_{p,k}\},\mathbf{p}_{p}, \forall p,\mathbf{L} } & \quad  \mathrm{Tr}\{\mathbf{W}\mathbf{L}^{-1}\}+\rho_1\sum_{p=1}^{P}\sum_{k=0}^{K} g(\mathbf{a}_{p,k}) \label{eq:binary integer constraint relaxation obj}\\
\mathrm{s.t. } &\quad  \left[ \begin{array}{ll}
\mathbf{F}_{\mathbf{ss}}-\mathbf{L} & \mathbf{F}_{\mathbf{s}\bm{\beta}} \\
\mathbf{F}_{\mathbf{s}\bm{\beta}}^T & \mathbf{F}_{\bm{\beta}\bm{\beta}}
\end{array}
\right]\succeq \mathbf{0}, \label{eq:binary integer constraint relaxation c1}\\
&\quad \mathbf{p}_{p}^T \mathbf{1}_{MN}\leq P_{\mathrm{tot}, p},\ \mathbf{p}_{p} \geq \mathbf{0}, \quad \forall p, \label{eq:binary integer constraint relaxation c2}\\
&\quad  \mathbf{0}_{MN} \preceq \mathbf{a}_{p,k} \preceq \mathbf{1}_{MN},\quad \forall p, k, \label{eq:binary integer constraint relaxation c3}\\
&\quad \sum_{p=1}^{P}\big(\mathbf{a}_{p,0}+\sum_{k=1}^{K}\mathbf{a}_{p,k}\big)\preceq \mathbf{1}_{MN}, \label{eq:binary integer constraint relaxation c6}\\
&\ R_\mathrm{c} \geq \eta_0, \label{eq:binary integer constraint relaxation c4}\\
&\ \max_{(l,\nu)\in\mathcal{S}} \big|\Gamma_{p}(l,\nu)\big|\leq \beta_0, \quad \forall p. \label{eq:binary integer constraint relaxation c5}
\end{align}
\end{subequations}
The term $\mathrm{Tr}\{\mathbf{W}\mathbf{L}^{-1}\}$ and all constraints are convex in $(\mathbf{L},\mathbf{p}_{p})$ for fixed $\{\mathbf{a}_{p,k}\}$, whereas the concave quadratic $g(\mathbf{a}_{p,k})$ renders problem~\eqref{eq:binary integer constraint relaxation} non-convex.
To circumvent this difficulty, we adopt a majorization--minimization (MM) approach, which iteratively constructs and minimizes a convex surrogate of the penalized objective.

Specifically, by applying the first-order Taylor expansion of $g(\mathbf{a}_{p,k})$ at the current iterate $\mathbf{a}_{p,k}^{(i)}$, we obtain the following affine upper bound:
\begin{subequations}\label{eq:MM relaxation}
\begin{align}
g(\mathbf{a}_{p,k})
&\leq \nabla g^T\big(\mathbf{a}_{p,k}^{(i)}\big)\big[\mathbf{a}_{p,k}-\mathbf{a}_{p,k}^{(i)}\big]
+ g\big(\mathbf{a}_{p,k}^{(i)}\big) \nonumber\\
&= \mathbf{a}^T_{p,k}\big[\mathbf{1}_{MN}-2\mathbf{a}_{p,k}^{(i)}\big]+ \|\mathbf{a}_{p,k}^{(i)}\|^2, \forall p, k, \label{eq:MM upper_bound}
\end{align}
\end{subequations}
where $\mathbf{a}_{p,k}^{(i)}$ denotes the value of $\mathbf{a}_{p,k}$ at the $i$-th iteration.

\subsubsection{Block Coordinate Descent Update}
 With the time--frequency selection variables $\{\mathbf{a}_{p,k}\}$ fixed, the subproblem of optimizing the power allocation vectors $\{\mathbf{p}_p\}$ reduces to
\begin{equation}\label{eq:update p}
\begin{aligned}
\min_{\mathbf{p}_{p}, \forall p, \mathbf{L}} & \quad  \mathrm{Tr}\{\mathbf{W}\mathbf{L}^{-1}\}\\
\mathrm{s.t. } &\quad  \eqref{eq:binary integer constraint relaxation c1}-
\eqref{eq:binary integer constraint relaxation c5},
\end{aligned}
\end{equation}
which is a convex optimization subproblem with linear matrix inequality  (LMI) constraints, and can be efficiently solved by standard solvers.

Given the updated power allocation $\mathbf{p}_{p}^{(i)}$ and the current iterate $\mathbf{a}_{p,k}^{(i)}$, the subproblem of updating the time--frequency selection variables $\mathbf{a}_{p,0}$ and $\mathbf{a}_{p,k}$ is
\begin{equation}\label{eq:update u}
\begin{aligned}
\min_{\{\mathbf{a}_{p,0}, \mathbf{a}_{p,k}\} \forall p,\, \mathbf{L}} & \quad  \mathrm{Tr}\{\mathbf{W}\mathbf{L}^{-1}\}
\!+\! \rho_1 \sum_{p=1}^{P}\sum_{k=0}^{K}\mathbf{a}^T_{p,k}\big[\mathbf{1}_{MN}-2\mathbf{a}_{p,k}^{(i)}\big]\\
\mathrm{s.t. } &\quad   \eqref{eq:binary integer constraint relaxation c1}, \eqref{eq:binary integer constraint relaxation c3}-\eqref{eq:binary integer constraint relaxation c5}.
\end{aligned}
\end{equation}
For fixed $(\mathbf{L},\mathbf{p}_{p}^{(i)})$, the surrogate objective in (\ref{eq:update u}) is affine in $\{\mathbf{a}_{p,k}\}$, while all constraints remain convex with respect to $\{\mathbf{a}_{p,k}\}$. Hence, this subproblem can also be efficiently solved.
By iterating the above two steps together with the MM update \eqref{eq:MM upper_bound}, the proposed algorithm monotonically decreases the penalized objective until convergence to a stationary point.
After convergence of the procedure, a final binary recovery step is applied: for each RE, if multiple relaxed variables are nonzero due to numerical tolerance, only the largest entry is retained and the others are set to zero; if all entries are below a small threshold, the RE is left unassigned.

\subsection{Summary and Computational Complexity Analysis}

Based on the above derivations, the proposed SLF based resource allocation design is summarized in Algorithm \ref{alg1}.
We now provide a brief complexity analysis.
Both subproblems \eqref{eq:update p} and \eqref{eq:update u} are convex optimization problems with additional LMI constraints and other convex constraints.
When an interior-point-based convex solver is used, the per-iteration complexity depends not only on the number of decision variables but also on the number and sizes of the LMI and other convex constraints.
In our formulation, the LMIs introduced by the Schur complement have small fixed sizes, while the dominant variable dimensions are $PMN$ for \eqref{eq:update p} and $P(K{+}1)MN$ for \eqref{eq:update u}.
Therefore, the overall scaling is dominated by the cubic terms, i.e., the computational complexity of solving \eqref{eq:update p} and \eqref{eq:update u} at each outer iteration is approximately
$\mathcal{O}\big((PMN)^3\big)$ and $\mathcal{O}\big((P(K{+}1)MN)^3\big)$, respectively.

\begin{table}[t]
\caption{Simulation Parameters}
\label{tab:system_settings}
\centering
\renewcommand{\arraystretch}{1.15}
\begin{tabular}{l c c}
\hline
\textbf{Parameter} & \textbf{Symbol} & \textbf{Value} \\
\hline
Carrier frequency            & $f_\mathrm{c}$            & $24\,\mathrm{GHz}$ \\
Subcarrier spacing           & $\Delta f$                & $120\,\mathrm{kHz}$ \\
Number of subcarriers        & $N$                       & $128$ \\
Number of symbols            & $M$                       & $64$ \\
Number of transmit BSs       & $P$                       & $2$ \\
Number of receive BSs        & $Q$                       & $2$ \\
Noise power                  & $\sigma^2$, $\sigma^2_\mathrm{comm}$ & $-80\, \mathrm{dBm}$ \\
Sum-rate threshold           & $\eta_0$                  & $5\, \mathrm{bps/Hz}$ \\
Number of users              & $K$                       & $2$ \\
Sidelobe-amplitude threshold                & $\beta_{0,\mathrm{dB}}$                 & $-10\, \mathrm{dB} $ \\
Range sidelobe regions       & $l_\mathrm{max}$          & $32$ \\
Velocity sidelobe regions    & $\nu_\mathrm{max}$        & $12$ \\
\hline
\end{tabular}
\end{table}

\begin{figure}[t]
 \centering
   \vspace{-0.0 cm}
  \includegraphics[width=  3.4  in]{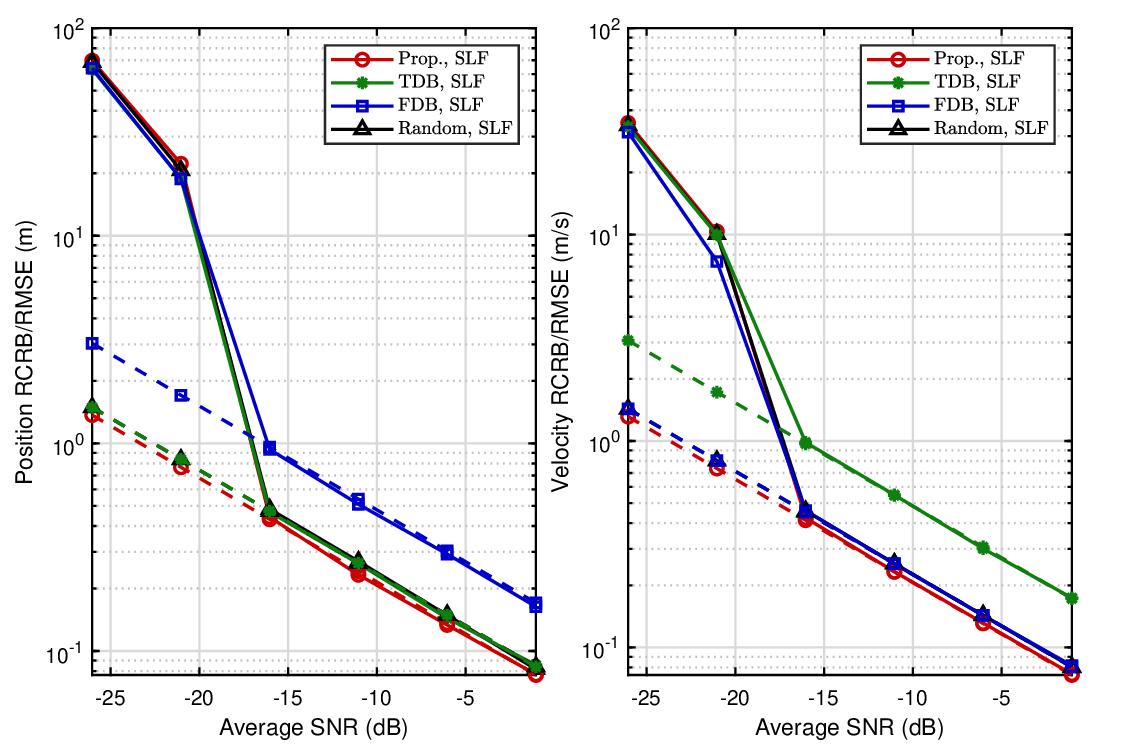}
  \caption{Position and velocity RMSE (solid) and square-root CRB (dashed) versus receive average SNR under SLF scheme.}\label{fig:RCRB_RMSE}
    \vspace{-0.2 cm}
\end{figure}

\section{Simulation Results}
In this section, we validate both the derived performance analysis and the proposed resource-allocation design. We consider two Tx-BSs located at $[-80,80]^T$ m and $[80,-80]^T$ m, and two Rx-BSs located at $[80,80]^T$ m and $[-80,-80]^T$ m. The target position and velocity are uniformly drawn from $[35,45]\times[30,40]$ m and $[10,20]\times[10,20]$ m/s, respectively. For communication, two downlink users are randomly generated within circles centered at $[-50,60]^T$ m and $[60,-45]^T$ m, each with radius $10$ m. The communication links follow Rayleigh fading with path loss $PL(d)=10^{(C_0/10)}d^{-\alpha}$, where $\alpha=2.4$ and $C_0=-30$ dB. Unless otherwise stated, we set $\lambda_p=\lambda_v=1$, $d_0=1$ m, and $v_0=1$ m/s, so that the weighting matrix reduces to $\mathbf{I}_4$. The other simulation parameters are summarized in Table~\ref{tab:system_settings}.

Fig.~\ref{fig:RCRB_RMSE} compares estimation performance of the proposed fine-grained 2D resource allocation design (``\textbf{Prop}.'') under SLF scheme. For comparison purposes, we also evaluate representative orthogonal baselines including time-division block allocation (``\textbf{TDB}''), frequency-division block allocation (``\textbf{FDB}''), and a random time--frequency allocation scheme (``\textbf{Random}'').
The average SNR is defined as the average receive SNR over all bistatic Tx–Rx links:
\begin{equation}
\mathrm{SNR}_\mathrm{avg}\triangleq 10\mathrm{log}_{10}\big(\frac{1}{PQ}\sum_{p=1}^{P}\sum_{q=1}^{Q}\frac{P_{\mathrm{tot},p}|\alpha_{p,q}|^2}{\sigma^2}\big).
\end{equation}
The proposed scheme consistently achieves the lowest position and velocity RMSEs by preserving the sensing aperture in both frequency and time, thereby retaining the effective bandwidth and CPI. In contrast, TDB sacrifices observation duration and thus degrades Doppler/velocity estimation, whereas FDB reduces bandwidth and therefore weakens position estimation. Although the random pattern partially preserves both spans, it cannot prioritize the most informative REs and thus remains inferior to the optimized 2D design. The close agreement between the RMSE curves and the corresponding CRBs also validates the CRB derivation and indicates that the SLF estimator operates near the performance bound.

\begin{figure}[t]
 \centering
   \vspace{-0.0 cm}
  \includegraphics[width=  3.4  in]{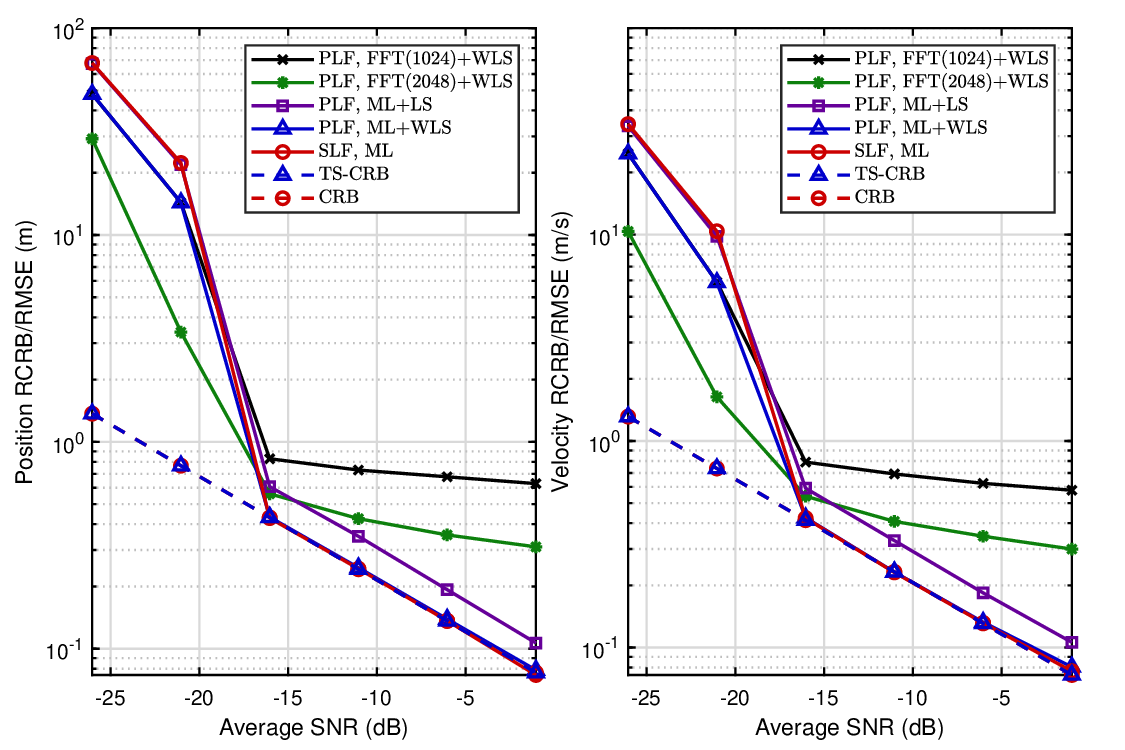}
  \caption{Position and velocity RMSE (solid) and square-root bounds (dashed) versus receive average SNR under PLF scheme.}\label{fig:PLF}
    \vspace{-0.2 cm}
\end{figure}

Fig.~\ref{fig:PLF} shows the performance loss due to local coarse estimation (using FFT instead of ML on local delay/Doppler extraction) and unweighted least-squares fusion in PLF scheme.
First, replacing continuous-parameter ML extraction with FFT-based processing introduces finite-grid compression before fusion, which causes grid mismatch, picket-fence loss, and finite-resolution quantization. Second, replacing covariance-aware WLS with plain LS introduces an additional fusion loss, because LS ignores the heterogeneous reliability of different bistatic links and therefore cannot fully exploit high-quality local estimates. Consequently, the oracle ML+WLS benchmark remains the closest PLF surrogate to the SLF bound. This agrees with the analysis in Section~III, which shows that PLF can asymptotically approach the SLF benchmark only under restrictive lossless conditions, including asymptotically efficient continuous-parameter ML extraction and covariance-aware WLS fusion. The PLF--SLF gap becomes less visible at low SNR, where noise dominates and the extra discretization-induced loss is comparatively less important.

\begin{figure}[t]
 \centering
   \vspace{-0.0 cm}
  \includegraphics[width=  \linewidth]{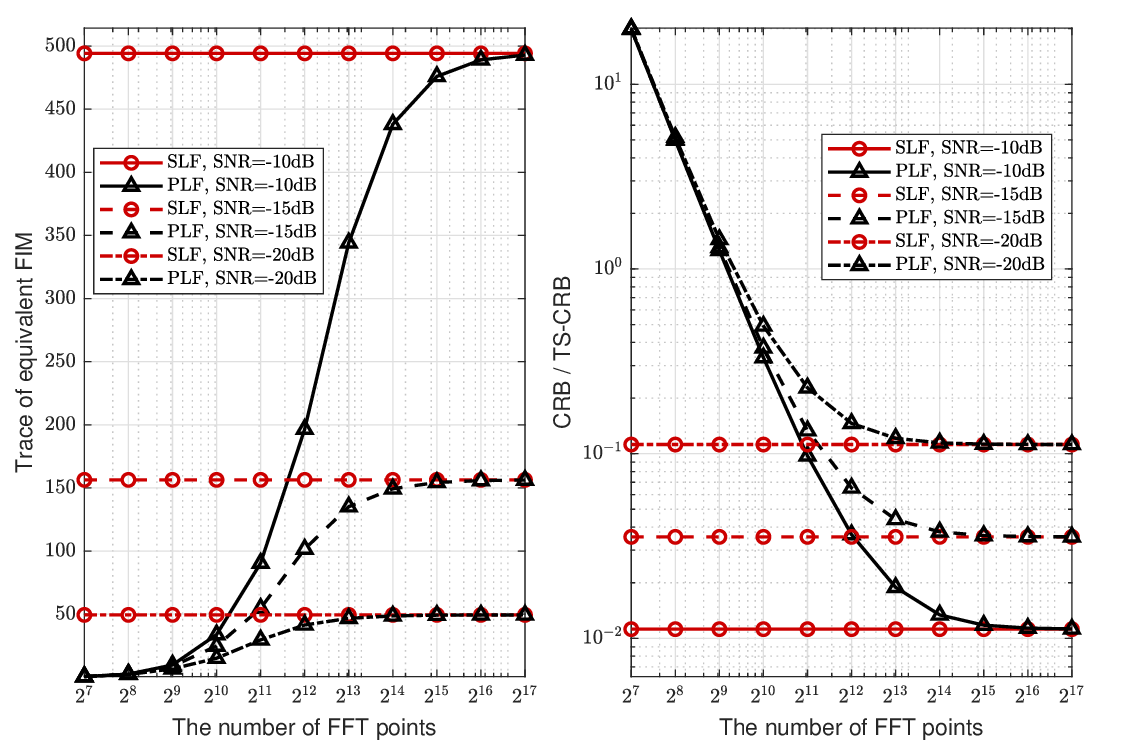}
  \caption{Effect of FFT resolution on PLF. }\label{fig:stage1}
    \vspace{-0.0 cm}
\end{figure}

Fig.~~\ref{fig:stage1} explains the behavior in Fig.~\ref{fig:PLF} by examining the impact of the FFT resolution on PLF.
The left figure shows the trace of equivalent Fisher-information $\mathrm{tr}\{\mathbf{F}_\mathrm{eqv}^\mathrm{PLF}\}$, and the right figure shows the corresponding CRB/TS-CRB.
As the number of FFT points per dimension in the 2D delay-doppler grid increases, the Fisher-information of PLF gradually approaches the SLF benchmark, while the corresponding TS-CRB monotonically decreases toward the CRB. This confirms that the performance loss of practical PLF originates from the finite-resolution local extractor. Another useful insight is that the convergence is faster at low SNR: when the operating point is noise-limited, the extra loss caused by a finite delay--Doppler grid becomes less dominant, so a moderate FFT size is already sufficient to approach the SLF benchmark. At higher SNR, however, the estimation error is increasingly dominated by grid-induced mismatch, and a much larger FFT size is required before the PLF performance saturates near the SLF limit.

\begin{figure}[t]
 \centering
   \vspace{-0.0 cm}
  \includegraphics[width=  \linewidth]{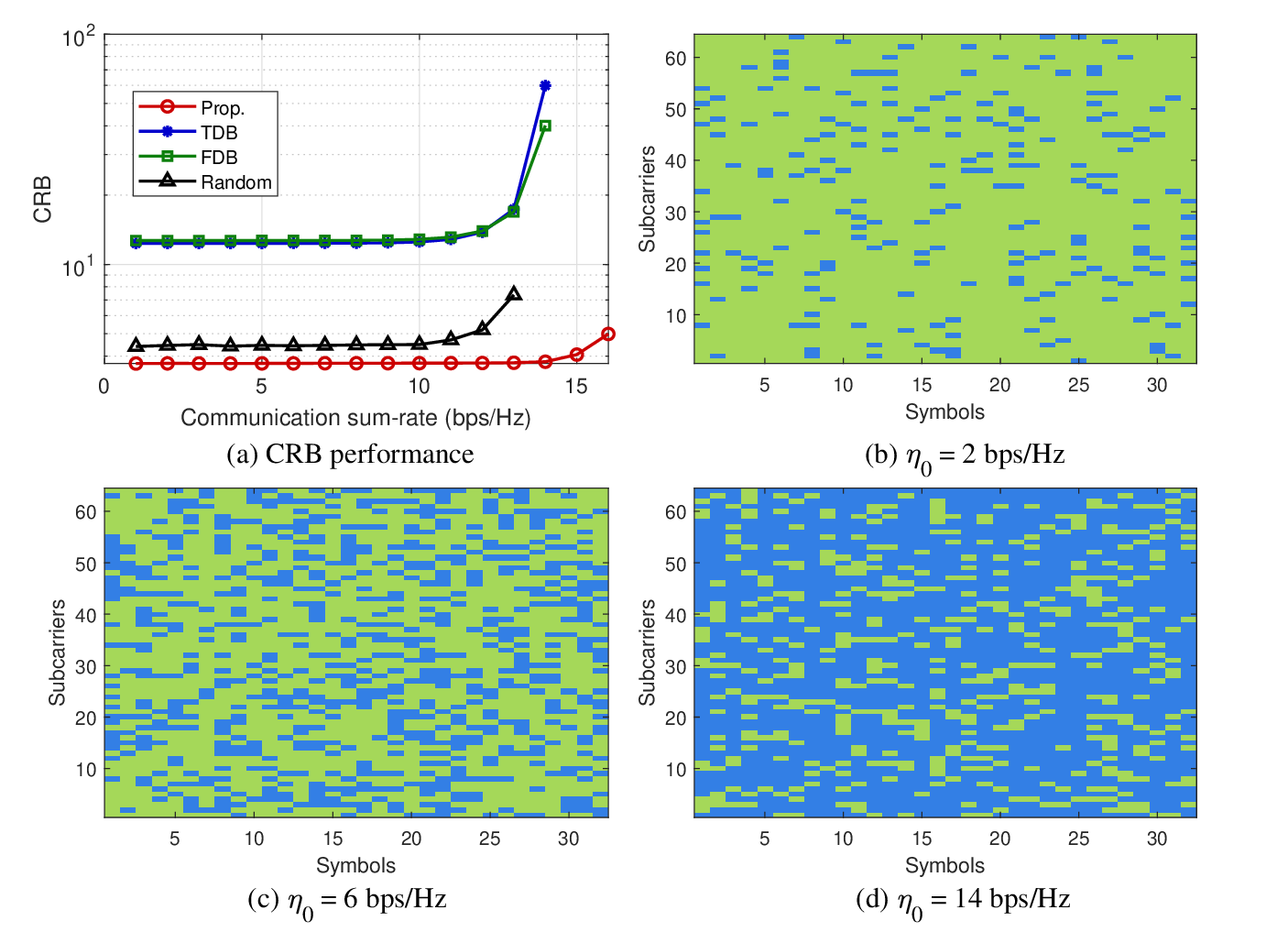}
  \caption{ Tradeoff between sensing accuracy and communication sum-rate. (a) CRB versus the required sum-rate. (b)--(d) Representative RE-allocation patterns of the proposed scheme at different rate requirements. (Sensing: \textcolor{color_radar}{\rule{7pt}{7pt}}, communication: \textcolor{color_comm}{\rule{7pt}{7pt}})
  }\label{fig:sumrate}
    \vspace{-0.2 cm}
\end{figure}

Fig.~\ref{fig:sumrate} illustrates the sensing--communication tradeoff. In Fig.~\ref{fig:sumrate}(a), the proposed scheme maintains a much lower sensing bound over a wide range of communication sum-rate requirements and degrades gracefully as $\eta_0$ increases. This behavior indicates that the optimized 2D allocation can first assign the most communication-efficient REs to downlink transmission while retaining the most sensing-informative REs, thereby postponing the onset of severe sensing degradation. By contrast, TDB and FDB already incur a large sensing penalty even at low communication loads because their fixed 1D partitions inherently destroy either the effective coherent processing interval or the effective bandwidth. The advantage of the proposed design is further visualized in Figs.~\ref{fig:sumrate}(b)-(d): as the sum-rate requirements increase, communication REs gradually occupy a larger fraction of the grid, while the remaining sensing REs of the proposed scheme stay distributed over both time and frequency rather than collapsing into a purely blockwise structure. In other words, the optimized design sacrifices sensing resources selectively rather than uniformly, which explains why it supports higher communication loads before entering the communication-limited regime.

\begin{figure}[t]
 \centering
   \vspace{-0.0 cm}
  \includegraphics[width= \linewidth]{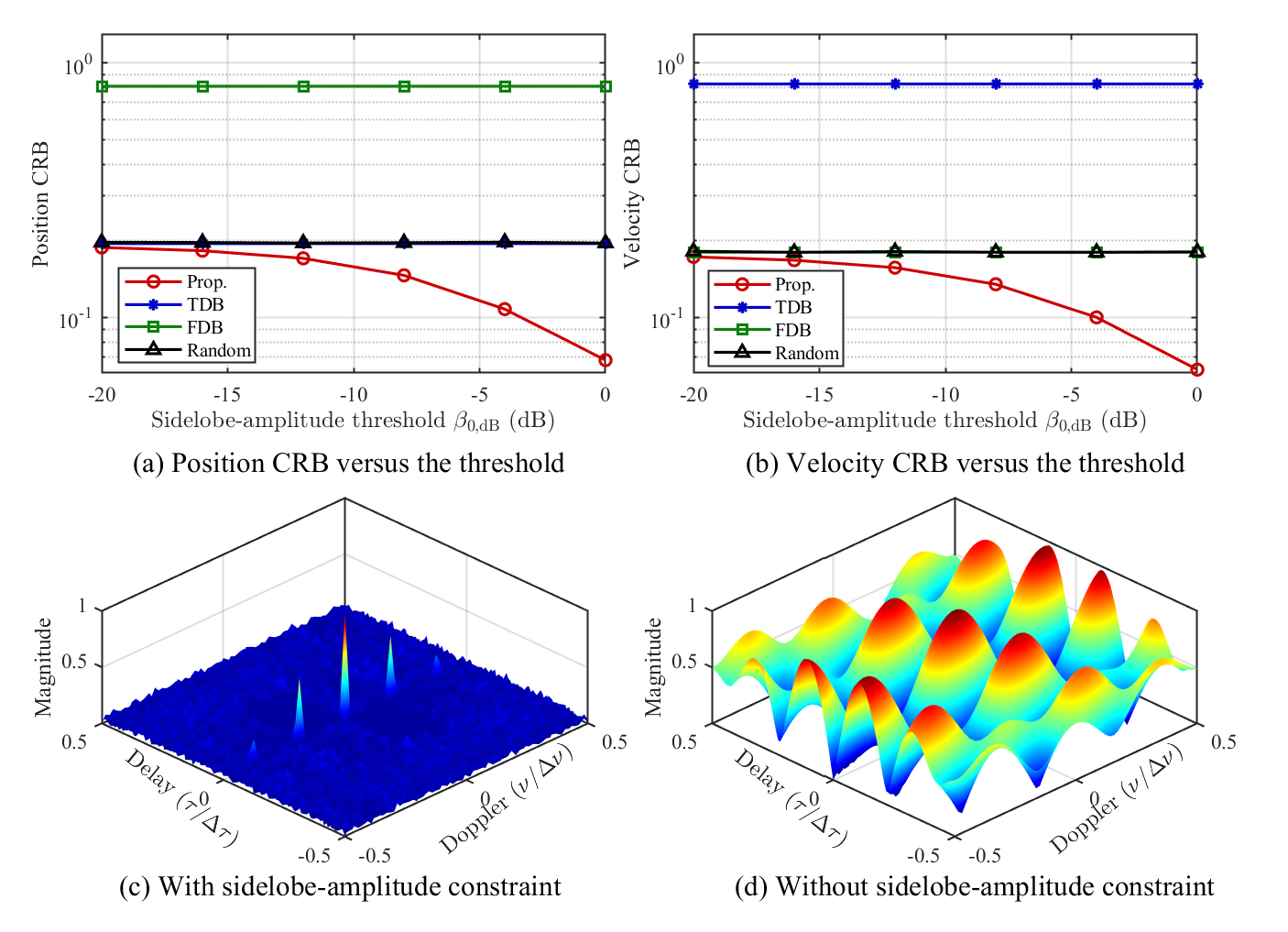}
  \caption{Effect of ambiguity-sidelobe control. (a) Position CRB versus the sidelobe-amplitude threshold.
   (b) Velocity CRB versus the sidelobe-amplitude threshold. (c) Ambiguity surface with the sidelobe-amplitude constraint. (d) Ambiguity surface without the sidelobe-amplitude constraint.}\label{fig:PSL necessity}
    \vspace{-0.2 cm}
\end{figure}

Fig.~\ref{fig:PSL necessity} demonstrates why ambiguity-sidelobe control is needed in addition to a CRB-driven objective. From Fig.~\ref{fig:PSL necessity}(a)-(b), relaxing the sidelobe-amplitude threshold slightly improves the sensing bound of the proposed scheme, but the gain is modest. This means that strong sidelobe suppression can be achieved at only a small loss in local bound optimality. The TDB and FDB baselines are much less sensitive to the sidelobe-amplitude threshold because their performance is mainly limited by reduced sensing span rather than by ambiguity shaping. The more important insight comes from Figs.~~\ref{fig:PSL necessity}(c) and ~\ref{fig:PSL necessity}(d): CRB-only optimization produces pronounced off-peak ambiguity ridges, whereas the sidelobe-amplitude-constrained design suppresses these sidelobes while preserving a sharp mainlobe. Hence, the sidelobe-amplitude constraint acts as a structural regularizer that complements the local CRB metric with global ambiguity robustness.

\begin{figure}[t]
 \centering
   \vspace{-0.0 cm}
  \includegraphics[width=  \linewidth]{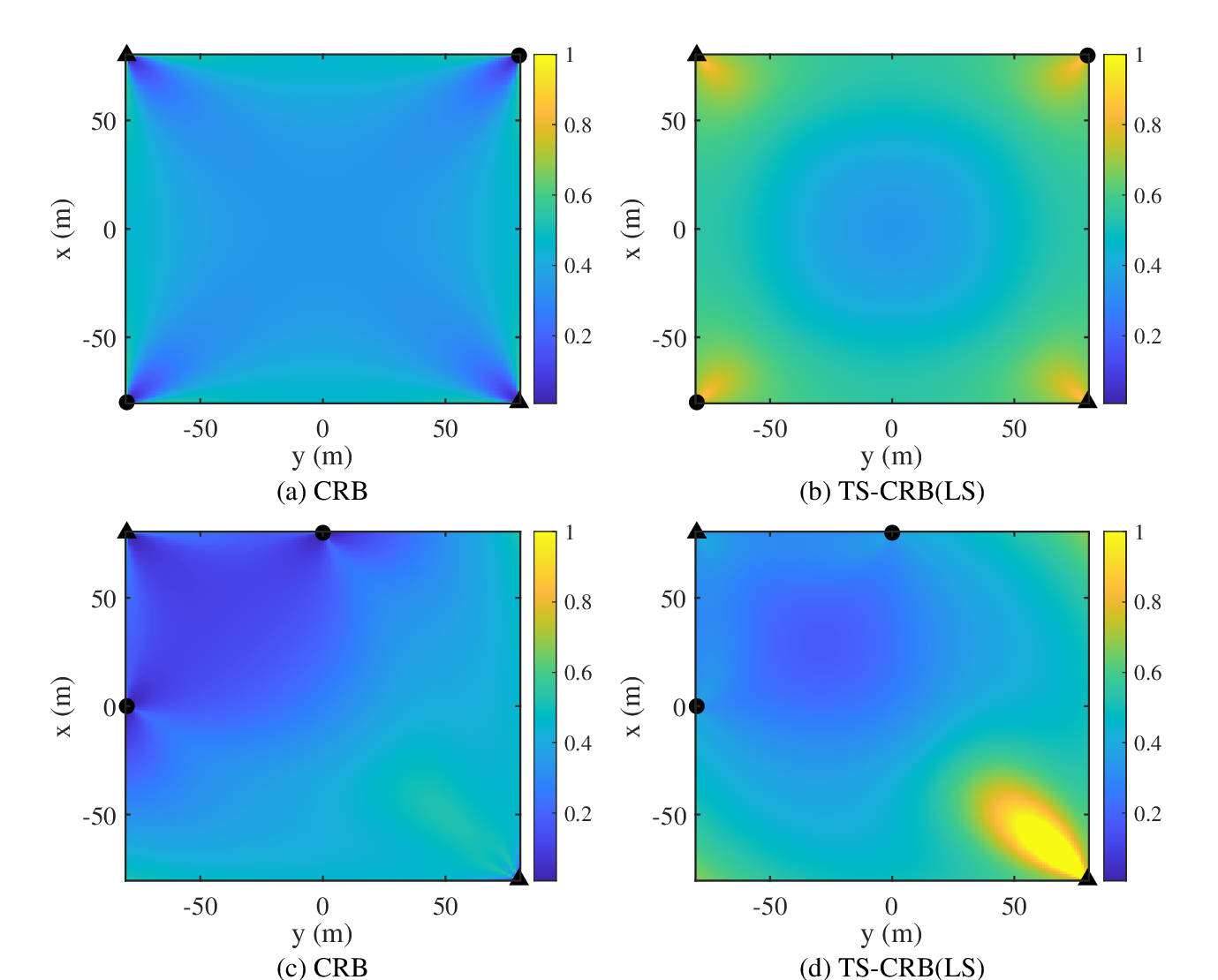}
  \caption{Spatial maps of the position estimation bound under SLF and PLF. (a) SLF with a symmetric deployment. (b) PLF with a symmetric deployment. (c) SLF with an asymmetric deployment. (d) PLF with an asymmetric deployment.
   (Triangle: Tx-BSs, circle: Rx-BSs)}\label{fig:PEB}
    \vspace{-0.2 cm}
\end{figure}

Fig.~\ref{fig:PEB} highlights the geometry dependence of the SLF--PLF gap under unweighted fusion, i.e., $\mathbf{W}_{\mathrm{WLS}}=\mathbf{I}_{2PQ}$. For the symmetric deployment in Figs.~\ref{fig:PEB} (a)-(b), PLF remains close to SLF over most of the central region because the bistatic links are relatively balanced and the geometric fusion problem is well conditioned. In contrast, for the asymmetric deployment in Figs.~\ref{fig:PEB} (c)-(d), PLF develops a pronounced high-bound region in poorly conditioned areas, whereas SLF remains much smoother over the entire surveillance region. This contrast shows that the performance loss of PLF is amplified when weak or geometrically unfavorable links are fused without proper weighting, because local delay/Doppler errors are directly propagated through the nonlinear geometry. The figure therefore provides a clear design insight: PLF is attractive in balanced deployments or fronthaul-limited regimes, whereas SLF is more robust in edge regions and asymmetric geometries.

\section{Conclusion}
This paper studied cooperative OFDM-ISAC networks from the dual perspectives of performance analysis and resource allocation. We derived the centralized CRB for SLF and a two-stage CRB-like metric for PLF, showing that SLF is the information-preserving benchmark, whereas only an oracle ML-based PLF can asymptotically approach it under restrictive conditions. Based on the SLF CRB, we developed a joint RE-selection and power-allocation design under communication, power, and sidelobe-amplitude constraints. Numerical results verified the analysis, showed clear gains of the proposed non-periodic 2D allocation over representative baselines, and further revealed that the SLF-PLF gap is mainly governed by local extraction fidelity, fusion weighting, and deployment geometry.

\appendices

\section{Derivation of the SLF CRB}\label{appendix:slf_crb}

Recall the vectorized SLF observation model in \eqref{eq:y_r_vector_revised}. For each bistatic link $(p,q)$, define
\begin{equation}\label{eq:app_gpq}
\mathbf g_{p,q}
\triangleq
\alpha_{p,q}
\big(
\bm{\psi}^{*}(f_{\mathrm d,p,q})
\otimes
\bm{\phi}(\tau_{p,q})
\big),
\end{equation}
so that the observation at the $q$-th Rx-BS can be written as
\begin{equation}\label{eq:app_yq_mu}
\mathbf y_q
=
\sum_{p=1}^{P}
\big(
\mathbf x_{p,0}\odot \mathbf g_{p,q}
\big)
+
\mathbf z_q.
\end{equation}
Stacking $\{\mathbf y_q\}_{q=1}^{Q}$ yields
\begin{equation}\label{eq:app_y_mu}
\mathbf y
=
\bm{\mu}(\bm{\eta})+\mathbf z,
\qquad
\mathbf z\sim\mathcal{CN}(\mathbf 0,\sigma^2\mathbf I).
\end{equation}

For the real-valued parameter vector \eqref{eq:define_eta},
the FIM entries satisfy
\begin{equation}\label{eq:app_fim_identity}
[\mathbf F_{\bm{\eta}}]_{i,j}
=
\frac{2}{\sigma^2}
\Re\!\left\{
\left(\frac{\partial \bm{\mu}}{\partial \eta_i}\right)^H
\left(\frac{\partial \bm{\mu}}{\partial \eta_j}\right)
\right\}.
\end{equation}

Due to the network-wide exclusivity of sensing REs, the pilot vectors of Tx-BSs have disjoint supports, i.e.,
\begin{equation}\label{eq:app_disjoint_support}
\mathbf x_{p,0}\odot \mathbf x_{p',0}=\mathbf 0,
\qquad p\neq p'.
\end{equation}
Hence, for a fixed Rx-BS $q$, derivative vectors associated with different Tx-BSs are orthogonal. Together with the independent stacking across different Rx-BSs, the global FIM admits the block form
\begin{equation}\label{eq:define_F_eta}
\mathbf F_{\bm{\eta}}
=
\begin{bmatrix}
\mathbf F_{1,1} & \mathbf F_{1,2} & \mathbf F_{1,3} & \mathbf F_{1,4}\\
\mathbf F_{2,1} & \mathbf F_{2,2} & \mathbf F_{2,3} & \mathbf F_{2,4}\\
\mathbf F_{3,1} & \mathbf F_{3,2} & \mathbf F_{3,3} & \mathbf F_{3,4}\\
\mathbf F_{4,1} & \mathbf F_{4,2} & \mathbf F_{4,3} & \mathbf F_{4,4}
\end{bmatrix},
\end{equation}
where each $\mathbf F_{i,j}\in\mathbb R^{PQ\times PQ}$ is diagonal, and the indices $i,j\in\{1,2,3,4\}$ correspond to
$\tau_{p,q}$, $f_{\mathrm d,p,q}$, $\Re\{\alpha_{p,q}\}$, and $\Im\{\alpha_{p,q}\}$, respectively.

For a fixed bistatic link $(p,q)$, define $\bm{\mu}_{p,q}
\triangleq
\mathbf x_{p,0}\odot \mathbf g_{p,q}$.
Then the derivative vectors are
\begin{equation}\label{eq:app_dmu_pq}
\frac{\partial \bm{\mu}_{p,q}}{\partial \eta_{p,q}^{(i)}}
=
\mathbf x_{p,0}\odot \dot{\mathbf g}_{p,q}^{(i)},
\qquad
\dot{\mathbf g}_{p,q}^{(i)}
\triangleq
\frac{\partial \mathbf g_{p,q}}{\partial \eta_{p,q}^{(i)}},
\quad i\in\{1,2,3,4\},
\end{equation}
and the corresponding diagonal entry of $\mathbf F_{i,j}$ is
\begin{equation}\label{eq:app_local_fim_entry}
[\mathbf F_{i,j}]_{r,r}
=
\frac{2}{\sigma^2}
\Re\!\left\{
\big(\mathbf x_{p,0}\odot \dot{\mathbf g}_{p,q}^{(i)}\big)^H
\big(\mathbf x_{p,0}\odot \dot{\mathbf g}_{p,q}^{(j)}\big)
\right\},
\end{equation}
where $r=(p-1)Q+q$.

Define the effective sensing power vector of Tx-BS $p$ as
\begin{equation}\label{eq:app_pp0_def}
\mathbf p_{p,0}
\triangleq
|\mathbf x_{p,0}|^{\odot 2}\in\mathbb R_{+}^{MN}.
\end{equation}
Under the sensing-only transmit model in (2) and unit-modulus pilots, we have $\mathbf p_{p,0}
=
\mathrm{vec}\big(\mathbf A_{p,0}\odot \mathbf P_p\big)$.
Using $(\mathbf x\odot \mathbf a)^H(\mathbf x\odot \mathbf b)
=
(|\mathbf x|^{\odot 2})^{T}(\mathbf a^{*}\odot \mathbf b)$,
\eqref{eq:app_local_fim_entry} becomes
\begin{equation}\label{eq:app_local_fim_linear}
[\mathbf F_{i,j}]_{r,r}
=
\frac{2}{\sigma^2}
\Re\!\left\{
\mathbf p_{p,0}^{T}
\widetilde{\mathbf g}_{p,q}^{(i,j)}
\right\},
\qquad
\widetilde{\mathbf g}_{p,q}^{(i,j)}
\triangleq
\big(\dot{\mathbf g}_{p,q}^{(i)}\big)^{*}
\odot
\dot{\mathbf g}_{p,q}^{(j)}.
\end{equation}
Therefore, every entry of $\mathbf F_{\bm{\eta}}$ is an explicit deterministic function of the sensing RE mask $\mathbf A_{p,0}$ and the power allocation matrix $\mathbf P_p$ through $\mathbf p_{p,0}$.

The required derivative vectors are
\begin{subequations}\label{eq:app_g_derivatives}
\begin{align}
\dot{\mathbf g}_{p,q}^{(1)}
&=
\frac{\partial \mathbf g_{p,q}}{\partial \tau_{p,q}}
=
\alpha_{p,q}
\big(
\bm{\psi}^{*}(f_{\mathrm d,p,q})\otimes \dot{\bm{\phi}}(\tau_{p,q})
\big),\\
\dot{\mathbf g}_{p,q}^{(2)}
&=
\frac{\partial \mathbf g_{p,q}}{\partial f_{\mathrm d,p,q}}
=
\alpha_{p,q}
\big(
\dot{\bm{\psi}}^{*}(f_{\mathrm d,p,q})\otimes \bm{\phi}(\tau_{p,q})
\big),\\
\dot{\mathbf g}_{p,q}^{(3)}
&=
\frac{\partial \mathbf g_{p,q}}{\partial \Re\{\alpha_{p,q}\}}
=
\bm{\psi}^{*}(f_{\mathrm d,p,q})\otimes \bm{\phi}(\tau_{p,q}),\\
\dot{\mathbf g}_{p,q}^{(4)}
&=
\frac{\partial \mathbf g_{p,q}}{\partial \Im\{\alpha_{p,q}\}}
=
\jmath\big(
\bm{\psi}^{*}(f_{\mathrm d,p,q})\otimes \bm{\phi}(\tau_{p,q})
\big),
\end{align}
\end{subequations}
where
\begin{subequations}\label{eq:app_phi_psi_derivatives}
\begin{align}
\dot{\bm{\phi}}(\tau)
&\triangleq
\frac{\partial \bm{\phi}(\tau)}{\partial \tau}
=
-\jmath 2\pi \Delta f\,
\bm{\varepsilon}_{N}\odot \bm{\phi}(\tau),\\
\dot{\bm{\psi}}(f_{\mathrm d})
&\triangleq
\frac{\partial \bm{\psi}(f_{\mathrm d})}{\partial f_{\mathrm d}}
=
-\jmath 2\pi T_{\mathrm s}\,
\bm{\varepsilon}_{M}\odot \bm{\psi}(f_{\mathrm d}),\\
\bm{\varepsilon}_{N}
&\triangleq
[0,1,\ldots,N-1]^{T},\\
\bm{\varepsilon}_{M}
&\triangleq
[0,1,\ldots,M-1]^{T}.
\end{align}
\end{subequations}

Recall the target-related parameter vector \eqref{eq:global_parameter_xi}
and the intermediate parameter vector \eqref{eq:intermediate_parameter_eta_pq}.
Since $\bm{\eta}$ is a differentiable function of $\bm{\xi}$, the chain rule yields
\begin{equation}\label{eq:app_chain_rule}
\mathbf F_{\bm{\xi}}
=
\mathbf J^{T}\mathbf F_{\bm{\eta}}\mathbf J,
\qquad
\mathbf J
\triangleq
\frac{\partial \bm{\eta}}{\partial \bm{\xi}^{T}}=\begin{bmatrix}
\mathbf J_\mathbf{s} & \mathbf 0\\
\mathbf 0 & \mathbf I_{2PQ}
\end{bmatrix},
\end{equation}

\begin{equation}\label{eq:app_J_s_total}
\mathbf J_\mathbf{s}
=
\frac{\partial \bm{\zeta}(\mathbf s)}{\partial \mathbf s^{T}}
=
\begin{bmatrix}
\mathbf J_{u} & \mathbf 0\\
\mathbf J_{u,v} & \mathbf J_{v}
\end{bmatrix},
\end{equation}
where
\begin{equation}
\begin{aligned}
\mathbf J_{u}\triangleq \frac{\partial \bm{\tau}}{\partial \mathbf u^{T}},
\
\mathbf J_{v}\triangleq \frac{\partial \mathbf f_{\mathrm D}}{\partial \mathbf v^{T}},
\
\mathbf{J}_{u,v}\triangleq \frac{\partial \mathbf f_{\mathrm D}}{\partial \mathbf u^{T}}
\in\mathbb R^{PQ\times 2}.
\end{aligned}
\end{equation}
In particular,
\begin{equation}\label{eq:app_Ju_Jv}
\mathbf J_{u}
\!=\!
\frac{1}{c_{0}}
\begin{bmatrix}
\mathbf r_{1,1}^{T}\\
\vdots\\
\mathbf r_{P,Q}^{T}
\end{bmatrix},
\quad
\mathbf J_{v}
\!=\!
\frac{1}{\lambda}
\begin{bmatrix}
\mathbf r_{1,1}^{T}\\
\vdots\\
\mathbf r_{P,Q}^{T}
\end{bmatrix},
\quad
\mathbf J_{u,v}
\!=\!
\begin{bmatrix}
\widehat{\mathbf{r}}^T_{1,1}\\
\vdots\\
\widehat{\mathbf{r}}^T_{P,Q}
\end{bmatrix},
\end{equation}
with
\begin{equation}\label{eq:app_rpq_def}
\mathbf r_{p,q}
=
\mathbf r_{p}+\mathbf r_{q},
\quad
\mathbf r_{p}
=
\frac{\mathbf u-\mathbf b_{p}}{\|\mathbf u-\mathbf b_{p}\|_{2}},
\quad
\mathbf r_{q}
=
\frac{\mathbf u-\mathbf b_{q}}{\|\mathbf u-\mathbf b_{q}\|_{2}},
\end{equation}
\begin{align}\label{eq:J_uv}
   \widehat{\mathbf{r}}^T_{p,q} = \frac{\mathbf{v}^T}{\lambda}\left(\frac{\mathbf{I}-\mathbf{r}_p\mathbf{r}_p^T}{\|\mathbf{u}-\mathbf{b}_p\|_2} + \frac{\mathbf{I}-\mathbf{r}_q\mathbf{r}_q^T}{\|\mathbf{u}-\mathbf{b}_q\|_2}\right).
\end{align}


\vspace{12pt}

\end{document}